\documentclass[preprint,prd,nofootinbib,tightenlines,groupedaddress,amsmath,amssymb]{revtex4}

\usepackage{bm}% bold math
\usepackage{color}
\usepackage{graphicx}% Include figure files
\usepackage[hypertex]{hyperref}
\usepackage{multirow}

\begin{document}
\baselineskip=15pt \parskip=3pt

\vspace*{3em}

\title{Probing Scotogenic Effects in $\bm{e^+e^-}$ Colliders}

\author{Shu-Yu Ho and Jusak Tandean}
\affiliation{Department of Physics and Center for Theoretical  Sciences, \\
National Taiwan University, \\ Taipei 106, Taiwan \vspace{7ex}}

%\date{\today}

%\pacs{13.66.Hk, 14.60.Pq, 14.80.Bn, 95.35.+d}

\begin{abstract}
We explore the possibility of employing $e^+e^-$ colliders to probe the scotogenic model,
in which neutrinos get mass radiatively via one-loop interactions involving dark matter.
Assuming the lightest one of the new particles in the model to be fermionic cold dark matter
and taking into account various constraints, including those from LHC Higgs experiments,
we show that LEP\,II data on $e^+e^-$ scattering into a pair of charged leptons plus missing
energy can place significant extra restrictions on the parameter space containing
sufficiently low masses of the charged scalars in the model.
On the other hand, LEP\,II data on $e^+e^-$ collisions into a photon plus missing energy
do not yield strong constraints.
The allowed parameter space can still accommodate Higgs exotic decays into the nonstandard
particles and thus is testable at the LHC.
We also consider using future measurements of these two types of $e^+e^-$ scattering at
the International Linear Collider to examine the scenario of interest further and find that
they can provide complementary information about it,
whether or not they reveal scotogenic effects.
\end{abstract}

\maketitle

\section{Introduction}

It goes without saying that the recent observation of a Higgs boson with mass around 126\,GeV
at the Large Hadron Collider (LHC)~\cite{lhc} and determination of the neutrino-mixing
parameter $\sin\theta_{13}$ at neutrino-oscillation experiments~\cite{An:2012eh} constitute
crucial guideposts for attempts to establish the nature of physics beyond the standard model~(SM).
Another factor that any realistic scenario for new physics would need to explain
is that about a quarter of the cosmic energy budget has been inferred from
astronomical observations to be attributable to dark matter~\cite{pdg,planck}.

One of the most economical possibilities accommodating the essential ingredients is
the scotogenic model invented by Ma~\cite{Ma:2006km}, in which neutrinos get
mass radiatively via one-loop interactions with nonstandard particles consisting of scalars
and fermions, at least one of which acts as dark matter~(DM).
Previously, within the context of this model we have
addressed~\cite{Ho:2013hia} some of the implications of the aforementioned experimental
findings, specifically the decays of the Higgs boson $h$ into final states containing
the new particles, assuming the lightest one of them to be fermionic cold DM.
Taking into account various experimental and theoretical constraints, we found that such
exotic decays of $h$ could have significant rates that were already probed by existing
LHC data and that the scotogenic effects on \,$h\to\gamma\gamma,\gamma Z$\, would be
testable in upcoming measurements.
In the present paper, we look at additional tests on this scenario of the model
using $e^+e^-$ colliders, motivated in part by the availability of good amounts of
past data from LEP\,II~\cite{lep:ee2llnn,lep:ee2gnn} that are potentially pertinent to our
parameter space of interest and in part by the increasing prospect of the International
Linear Collider~(ILC) being realized in the foreseeable future~\cite{ilc}.

The structure of the paper is as follows.
In the next section, we first describe the relevant Lagrangians for the nonstandard particles
in the model and the expressions related to the neutrino masses.
Subsequently, adopting the Particle Data Group (PDG) parametrization of the neutrino-mixing
matrix, we derive exact solutions for the Yukawa couplings of the new particles in terms of
only three free parameters.
We will pick one set of such solutions to be used in our numerical work.
In~Section~\ref{constraints}, we briefly review the main restraints on the parameter space
under consideration and also employ the Planck data on the DM relic
abundance to update the allowed ranges of the Yukawa coupling belonging to the DM candidate.
In~Section~\ref{colliders}, with the parameter values satisfying the constraints listed
earlier, we investigate the scotogenic effects on the Higgs boson decay, taking into account
other restrictions from the latest LHC data.
Moreover, we explore complementary and further tests on the model from past measurements
on $e^+e^-$ collisions at LEP\,II.
In particular, we show that the LEP\,II data on $e^+e^-$ scattering into a pair of charged
leptons plus missing energy can impose potentially important extra constraints, much more so
than the data on $e^+e^-$ colliding into a~photon plus missing energy.
Nevertheless, we also find that experiments on the two types of $e^+e^-$ scattering processes
at the future ILC can supply complementary results useful for probing the model.
We give our conclusions in~Section~\ref{summary}.
Some additional information and lengthy formulas are collected in a~couple of appendixes.

\section{Interactions and Yukawa couplings\label{interactions}}

In the simplest version of the scotogenic model~\cite{Ma:2006km,Bouchand:2012dx},
the components beyond the minimal SM are a~scalar doublet,~$\eta$, and three singlet Majorana
fermions, $N_{1,2,3}$, all of which are odd under an~exactly conserved $Z_2$ symmetry.
All of the SM particles are $Z_2$ even.
It follows that the lightest one of the nonstandard particles is stable and can serve as~DM.
Here we suppose that $N_1$ is a good candidate for cold~DM.

The Lagrangian responsible for the interactions of the scalar particles in this model with
one another and with the gauge bosons is
\begin{eqnarray} \label{L}
{\cal L} \,\,=\,\, ({\cal D}^\rho\Phi)^\dagger\,{\cal D}_\rho^{}\Phi \,+\,
({\cal D}^\rho\eta)^\dagger\,{\cal D}_\rho^{}\eta \;-\; {\cal V} ~,
\end{eqnarray}
where ${\cal D}_\rho$ denotes the usual covariant derivative containing the SM gauge fields,
the potential~\cite{Ma:2006km}
\begin{eqnarray} \label{potential}
{\cal V} &\,=&\, \mu_1^2\,\Phi^\dagger \Phi \,+\, \mu_2^2\,\eta^\dagger\eta \,+\,
\mbox{$\frac{1}{2}$}\lambda_1^{}(\Phi^\dagger \Phi)^2 \,+\,
\mbox{$\frac{1}{2}$}\lambda_2^{}(\eta^\dagger\eta)^2
\nonumber \\ && +\;
\lambda_3^{}(\Phi^\dagger \Phi)(\eta^\dagger\eta) \,+\,
\lambda_4^{}(\Phi^\dagger\eta)(\eta^\dagger \Phi) \,+\,
\mbox{$\frac{1}{2}$}\lambda_5^{}\bigl[ (\Phi^\dagger\eta)^2+(\eta^\dagger\Phi)^2\bigr] ~,
\end{eqnarray}
and, after electroweak symmetry breaking,
\begin{eqnarray}
\Phi \,\,=\, \left(\!\begin{array}{c} 0 \vspace{1pt} \\
\frac{1}{\sqrt2}_{\vphantom{o}}(h+v) \end{array}\! \right) , \hspace{7ex}
\eta \,\,=\, \left(\!\begin{array}{c} H^+ \vspace{1pt} \\
\frac{1}{\sqrt2}_{\vphantom{o}}({\cal S}+i{\cal P}) \end{array}\! \right) ,
\end{eqnarray}
with $h$ being the physical Higgs boson and $v$ the vacuum expectation value (VEV) of $\Phi$.
The $Z_2$ symmetry implies that the VEV of $\eta$ is zero.
The masses of $\cal S$, $\cal P$, and $H^\pm$ are then given by
\begin{eqnarray}
m_{\cal S}^2 \,\,=\,\, m_{\cal P}^2 \,+\, \lambda_{5\,}v^2  \,\,=\,\,
\mu_2^2 \,+\, \mbox{$\frac{1}{2}$}(\lambda_3+\lambda_4+\lambda_5)v^2 ~, \hspace{7ex}
m_H^2 \,\,=\,\, \mu_2^2 \,+\, \mbox{$\frac{1}{2}$}\lambda_3\,v^2 ~.
\end{eqnarray}
We work under the assumption that $\lambda_5$ is very small~\cite{Kubo:2006yx},
\,$|\lambda_5|\ll|\lambda_3+\lambda_4|$,\, implying that $m_{\cal S,P}$ are nearly degenerate,
\,$|m_{\cal S}^2-m_{\cal P}^2|=|\lambda_5|v^2\ll m_{\cal S}^2\simeq m_{\cal P}^2$.\,
In $\cal L$, the part that includes the couplings of $\eta$ to $h$, the photon~$A$, and
the $Z$ boson is
\begin{eqnarray} \label{Lphieta}
{\cal L} &\,\supset&\, \bigl[ \bigl(\mu_2^2-m_{\cal S}^2\bigr){\cal S}^2 +
\bigl(\mu_2^2-m_{\cal P}^2\bigr){\cal P}^2 + 2\bigl(\mu_2^2-m_H^2\bigr)H^+H^- \bigr] \frac{h}{v}
\nonumber \\ && +\;
i e\,\bigl(H^+\,\partial^\rho H^--H^-\,\partial^\rho H^+\bigr) A_\rho^{}
\,+\, e^2\,H^+H^- A^2
\,+\, \frac{e g}{c_{\rm w}}\bigl(1-2s_{\rm w}^2\bigr)\,H^+H^- A^\rho Z_\rho^{}
\nonumber \\ && +\;
\frac{g}{2c_{\rm w}} \bigl[ {\cal P}\,\partial^\rho{\cal S}-{\cal S}\,\partial^\rho{\cal P}
\,+\, i\bigl(1-2s_{\rm w}^2\bigr) \bigl(H^+\,\partial^\rho H^--H^-\,\partial^\rho H^+\bigr)
\bigr] Z_\rho^{} ~,
\end{eqnarray}
where only terms pertinent to the processes we discuss are on display, \,$e=g s_{\rm w}^{}>0$\,
is the electromagnetic charge, and
\,$c_{\rm w}^{}=\bigl(1-s_{\rm w}^2\bigr)\raisebox{0.7pt}{$^{1/2}$}=\cos\theta_{\rm W}$\,
with the Weinberg angle $\theta_{\rm W}$.

The Lagrangian for the masses and interactions of the new singlet fermions $N_k$ is
\begin{eqnarray} \label{LN}
{\cal L}_N^{} \,\,=\,\,
-\mbox{$\frac{1}{2}$} M_k^{}\,\overline{N_k^{\rm c}}\,P_R^{} N_k^{} \,+\,
{\cal Y}_{rk}^{} \Bigl[ \bar\ell_r^{} H^- \,-\,
\mbox{$\frac{1}{\sqrt2}$}\,\bar\nu_r^{}\,({\cal S}-i {\cal P}) \Bigr] P_R^{} N_k^{}
\;+\; {\rm H.c.} ~,
\end{eqnarray}
where $M_k$ denote their masses, summation over \,$k,r=1,2,3$\, is implicit,
the superscript c refers to charge conjugation, \,$P_R=\frac{1}{2}(1+\gamma_5)$,\,
and \,$\ell_{1,2,3}=e,\mu,\tau$.\,
The Yukawa couplings of $N_k$ make up the matrix
\begin{eqnarray} \label{yukawa}
{\cal Y} \,\,=\, \left(\begin{array}{ccc} Y_{e1} & Y_{e2} & Y_{e3} \vspace{2pt} \\
Y_{\mu1} & Y_{\mu2} & Y_{\mu3} \vspace{2pt} \\ Y_{\tau1} & Y_{\tau2} & Y_{\tau3}
\end{array}\right) ,
\end{eqnarray}
where \,$Y_{\ell_r k}={\cal Y}_{r k}^{}$.\,

\newpage

The light neutrinos acquire mass radiatively through one-loop diagrams with internal
$\cal S$, $\cal P$, and~$N_k$.
The resulting mass eigenvalues $m_j^{}$ are given by~\cite{Ma:2006km}
\begin{eqnarray} \label{UMU} & \displaystyle
{\rm diag}\bigl(m_1^{},m_2^{},m_3^{}\bigr) \,\,=\,\, {\cal U}^\dagger{\cal M}_\nu\,{\cal U}^* \,,
& \\ \label{Mnu} & \displaystyle
{\cal M}_\nu^{} \,\,=\,\, {\cal Y}\, {\rm diag}(\Lambda_1,\Lambda_2,\Lambda_3)\,
{\cal Y}^{{\rm T}^{\vphantom{\displaystyle|}}}  \,,
& \\ \label{lambdak} & \displaystyle
\Lambda_k^{} \,\,=\,\, \frac{\lambda_{5\,}^{}v^2}{16\pi^2M_k^{}}\Biggl[ \frac{M_k^2}{m_0^2-M_k^2}
+ \frac{2 M_k^4\,\ln\bigl(M_k^{}/m_0^{}\bigr)}{\bigl(m_0^2-M_k^2\bigr)\raisebox{1.7pt}{$^{\!2}$}}
\Biggr]^{\vphantom{\int^|}} \,, \hspace{7ex}
m_0^2 \,\,=\,\, \mbox{$\frac{1}{2}$}\bigl(m_{\cal S}^2 + m_{\cal P}^2\bigr) ~, &
\end{eqnarray}
where $\,\cal U$ is the Pontecorvo-Maki-Nakagawa-Sakata (PMNS~\cite{pmns}) unitary matrix and
the formula for $\Lambda_k^{}$ applies to
the \,$m_0^{}\simeq m_{\cal S}^{}\simeq m_{\cal P}^{}$\, case.

For the $\cal U$ matrix we choose the PDG parametrization~\cite{pdg}
\begin{eqnarray} \label{U} & \displaystyle
{\cal U} \,\,=\,\, \tilde u\;{\rm diag}\bigl(e^{i\alpha_1/2},e^{i\alpha_2/2},1\bigr) ~,
& \\ & \displaystyle
\tilde u \,\,=\, \left(\begin{array}{ccc}
 c_{12\,}^{}c_{13}^{} & s_{12\,}^{}c_{13}^{} & s_{13}^{}\,e^{-i\delta^{\vphantom{|}}}
\vspace{3pt} \\
-s_{12\,}^{}c_{23}^{}-c_{12\,}^{}s_{23\,}^{}s_{13}^{}\,e^{i\delta} & ~~
 c_{12\,}^{}c_{23}^{}-s_{12\,}^{}s_{23\,}^{}s_{13}^{}\,e^{i\delta} ~~ & s_{23\,}^{}c_{13}^{}
\vspace{3pt} \\
 s_{12\,}^{}s_{23}^{}-c_{12\,}^{}c_{23\,}^{}s_{13}^{}\,e^{i\delta} &
-c_{12\,}^{}s_{23}^{}-s_{12\,}^{}c_{23\,}^{}s_{13}^{}\,e^{i\delta} & c_{23\,}^{}c_{13}^{}
\end{array}\right)^{\vphantom{\int^|}} , &
\end{eqnarray}
where \,$\delta\in[0,2\pi]$\, and \,$\alpha_{1,2}^{}\in[0,2\pi]$\, are the Dirac and Majorana
$CP$-violation phases, respectively, \,$c_{mn}^{}=\cos\theta_{mn}^{}\ge0$,\, and
\,$s_{mn}^{}=\sin\theta_{mn}^{}\ge0$.\,
A recent analysis of global neutrino-oscillation data
yields~\cite{GonzalezGarcia:2012sz}\footnote{Somewhat earlier analyses of the global neutrino
data in Ref.\,\cite{Tortola:2012te} produced similar results.}
\begin{eqnarray} \label{xranges}
s_{12}^2 \,=\, 0.302_{-0.012}^{+0.013} ~, ~~~~~~~
s_{23}^2 \,=\, 0.413_{-0.025}^{+0.037} ~, ~~~~~~~
s_{13}^2 \,=\, 0.0227_{-0.0024}^{+0.0023} ~, ~~~~~~~
\delta \,=\, \bigl(300_{-138}^{+66}\bigr)^\circ \,. ~~
\end{eqnarray}

Upon applying Eq.\,(\ref{U}) in Eq.\,(\ref{UMU}), we arrive at the relations
\begin{eqnarray} \label{numass} & \displaystyle
m_r^{} \,\,=\,\, e^{-i\alpha_r^{}}\raisebox{2pt}{\footnotesize$\displaystyle\sum_k$}\,
X_{rk\,}^2\Lambda_k^{} ~, \hspace{7ex}
\raisebox{2pt}{\footnotesize$\displaystyle\sum_k$}\, X_{rk\,}^{}X_{ok\,}^{}\Lambda_k^{} \,\,=\,\, 0 ~,
& \\ & \displaystyle
X_{rk}^{} \,\,=\,\, \bigl(\tilde u^\dagger{\cal Y}\bigr)_{rk} ~, \hspace{7ex}
\alpha_3^{} \,\,=\,\, 0 ~, \hspace{7ex} k,o,r \,\,=\,\, 1,2,3 ~, ~~~~ o\,\,\neq\,\,r ~. &
\end{eqnarray}
Explicitly,
\begin{eqnarray}
X_{1k}^{} &\,=\,& c_{12\,}^{}c_{13\,}^{}Y_{ek}^{}
\,-\, \Bigl(s_{12\,}^{}c_{23}^{}+c_{12\,}^{}s_{23\,}^{}s_{13\,}^{}e^{-i\delta}\Bigr)Y_{\mu k}^{}
\,+\, \Bigl(s_{12\,}^{}s_{23}^{}-c_{12\,}^{}c_{23\,}^{}s_{13\,}^{}e^{-i\delta}\Bigr)Y_{\tau k}^{} ~,
\nonumber \\
X_{2k}^{} &\,=\,& s_{12\,}^{}c_{13\,}^{}Y_{ek}^{}
\,+\, \Bigl(c_{12\,}^{}c_{23}^{}-s_{12\,}^{}s_{23\,}^{}s_{13\,}^{}e^{-i\delta}\Bigr)Y_{\mu k}^{}
\,-\, \Bigl(c_{12\,}^{}s_{23}^{}+s_{12\,}^{}c_{23\,}^{}s_{13\,}^{}e^{-i\delta}\Bigr)Y_{\tau k}^{} ~,
\nonumber \\
X_{3k} &\,=\,& s_{13\,}^{}e^{i\delta\,}Y_{ek}^{} \,+\, s_{23\,}^{}c_{13\,}^{}Y_{\mu k}^{}
\,+\, c_{23\,}^{}c_{13\,}^{}Y_{\tau k}^{} ~.
\end{eqnarray}
The diagonalization conditions in Eq.\,(\ref{numass}) turn out to be exactly solvable for
two of the three elements $Y_{\ell_r k}$ with the same $k$ in terms of the third one,
in which case the $\cal Y$ matrix has only three free (complex) parameters.
We opt for getting $Y_{e k}$ and $Y_{\mu k}$ in terms of \,$Y_k\equiv Y_{\tau k}$.\,
As outlined in Appendix\,\,\ref{yjk}, there is more than one set of the solutions,
but not all of the sets fulfill the requirement that at least two of
the mass eigenvalues $m_{1,2,3}^{}$ be nonzero.

One of the solution sets that can supply three nonzero masses of the neutrinos comprises
\begin{eqnarray} \label{123} & \displaystyle
Y_{e1}^{} \,\,=\,\, \frac{-c_{12\,}^{}c_{13}^{}\,Y_1^{}}
{c_{12\,}^{}c_{23\,}^{}s_{13\,}^{}e^{i\delta}-s_{12\,}^{}s_{23}^{}} ~, ~~~~~~~ &
Y_{\mu1}^{} \,\,=\,\, \frac{c_{12\,}^{}s_{23\,}^{}s_{13\,}^{}e^{i\delta}+s_{12\,}^{}c_{23}^{}}
{c_{12\,}^{}c_{23\,}^{}s_{13\,}^{}e^{i\delta}-s_{12\,}^{}s_{23}^{}}\; Y_1^{} ~,
\nonumber \\ & \displaystyle
Y_{e2}^{} \,\,=\,\, \frac{-s_{12\,}^{}c_{13}^{}\,Y_2^{\vphantom{\int}}}
{s_{12\,}^{}c_{23\,}^{}s_{13\,}^{}e^{i\delta}+c_{12\,}^{}s_{23}^{}} ~, ~~~~~~~ &
Y_{\mu2}^{} \,\,=\,\, \frac{s_{12\,}^{}s_{23\,}^{}s_{13\,}^{}e^{i\delta}-c_{12\,}^{}c_{23}^{}}
{s_{12\,}^{}c_{23\,}^{}s_{13\,}^{}e^{i\delta}+c_{12\,}^{}s_{23}^{}}\; Y_2^{} ~,
\nonumber \\ & \displaystyle
Y_{e3}^{} \,\,=\,\, \frac{s_{13}^{}\,Y_3^{}}{c_{23\,}^{}c_{13\,}^{}e^{i\delta}} ~, \hspace{17ex} &
Y_{\mu3}^{} \,\,=\,\, \frac{s_{23\,}^{}Y_3^{\vphantom{\int}}}{c_{23}^{}} ~,
\end{eqnarray}
which correspond to the mass eigenvalues
\begin{eqnarray} \label{m1m2m3}
m_1^{} \,=\, \frac{\Lambda_{1\,}^{}Y_{e1}^2\,e^{-i\alpha_1}}{c_{12\,}^2c_{13}^2} ~, \hspace{7ex}
m_2^{} \,=\, \frac{\Lambda_{2\,}^{}Y_{e2}^2\,e^{-i\alpha_2}}{s_{12\,}^2c_{13}^2} ~, \hspace{7ex}
m_3^{} \,=\, \frac{\Lambda_{3\,}^{}Y_3^2}{c_{13\,}^2 c_{23}^2} ~.
\end{eqnarray}
The necessity that $m_{1,2,3}^{}$ be real and nonnegative then implies that
\begin{eqnarray} & \displaystyle
\alpha_1^{} \,\,=\,\, \arg\bigl(\Lambda_{1\,}^{}Y_{e1}^2\bigr) ~, \hspace{7ex}
\alpha_2^{} \,\,=\,\, \arg\bigl(\Lambda_{2\,}^{}Y_{e2}^2\bigr) ~, \hspace{7ex}
\arg\bigl(\Lambda_{3\,}^{}Y_3^2\bigr) \,\,=\,\, 0 ~.
\end{eqnarray}
In the rest of the paper, we utilize Eqs.\,\,(\ref{123}) and (\ref{m1m2m3}),
and for simplicity we set \,$e^{i\delta}=1$,\, in accord with
the empirical range of $\delta$ in~Eq.\,(\ref{xranges}).
Also, we take $Y_{1,2,3}$ to be real and nonnegative.

Now, in our previous study we adopted a simpler form of $\cal U$ which depends on only two
angles, $\theta$ and $\varsigma$, and has no phases~\cite{Ho:2013hia}.
It can be reproduced from $\tilde u$ in Eq.\,(\ref{U}) with
\begin{eqnarray} \label{sc}
s_{12}^{} \,\,=\,\, \frac{s_\theta^{}}{\sqrt{1-c_{\theta\,}^2 s_\varsigma^2}} ~, \hspace{7ex}
s_{23}^{} \,\,=\,\, \frac{c_\varsigma^{}-s_{\theta\,}^{}s_\varsigma^{}}
{\sqrt{2-2c_{\theta\,}^2 s_\varsigma^2}} ~, \hspace{7ex}
s_{13}^{} \,\,=\,\, c_{\theta\,}^{}s_\varsigma^{} ~,
\end{eqnarray}
and \,$\delta=0$,\, where \,$c_a^{}=\cos a$\, and \,$s_a^{}=\sin a$.\,
Moreover, numerically we chose for definiteness \,$\theta=32.89^\circ$\, and
\,$c_{\theta\,}^{}s_{\varsigma}^{}=\sqrt{0.0227}$,\,
which led to $\,\cal U$ elements in agreement at the one-sigma
level with their experimental values in~Eq.\,(\ref{xranges}).
Hence, in the present analysis we adopt for $\tilde u$ the same numerical input.
According to Eq.\,(\ref{sc}), this translates into \,$s_{12}^2\simeq 0.302$,\,
\,$s_{23}^2\simeq 0.402$,\, and~\,$s_{13}^2=0.0227$,\, consistent with
Eq.\,(\ref{xranges}) and leading to the neutrino
eigenmasses\footnote{It is instructive to see how $m_{1,2,3}$ would be modified with
a~tribimaximal form~\cite{tribi} of the mixing matrix $\,\cal U$, which corresponds to
\,$\bigl(s_{12}^2,s_{23}^2,s_{13}^2\bigr)=(1/3,1/2,0)$\,
and is therefore no longer compatible with the current data~\cite{GonzalezGarcia:2012sz}.
Applying this to Eq.\,(\ref{m1m2m3}), with zero phases, yields
\,$\bigl(m_1^{},m_2^{},m_3^{}\bigr)=\bigl(6\Lambda_1^{}Y_1^2,\,3\Lambda_2^{}Y_2^2,\,
2\Lambda_3^{}Y_3^2\bigr)$\,~\cite{Kashiwase:2013uy}.
Hence the $m_1^{}$ value is very different from that in Eq.\,(\ref{m123}).\vspace{-2ex}}
\begin{eqnarray} \label{m123}
m_1^{} \,\,\simeq\,\, 15.9_{\,}\Lambda_1^{}Y_1^2 ~, \hspace{7ex}
m_2^{} \,\,\simeq\,\, 2.8_{\,}\Lambda_2^{}Y_2^2 ~, \hspace{7ex}
m_2^{} \,\,\simeq\,\, 1.7_{\,}\Lambda_3^{}Y_3^2 ~.
\end{eqnarray}

\section{Constraints from low energy and DM data\label{constraints}}

As we discussed in Ref.\,\cite{Ho:2013hia}, there are a number of theoretical and
experimental restrictions on the couplings and masses of the nonstandard particles in
the scotogenic scenario being examined.
We found specifically that the strictest limitations on the Yukawa couplings
${\cal Y}_{rk}$ come from the data on the neutrino squared-mass differences
\,$\Delta_{jk}^2=m_j^2-m_k^2$,\, the empirical bounds on the branching ratios of
the charged-lepton flavor-changing radiative decays \,$\ell_i\to\ell_j\gamma$,\,
and the measurement of the muon anomalous magnetic moment $a_\mu^{}$.
The parameter space of interest in this study is subject to the same restraints
from low-energy experiments.

\begin{figure}[ht] \vspace{7pt}
\includegraphics[width=86mm]{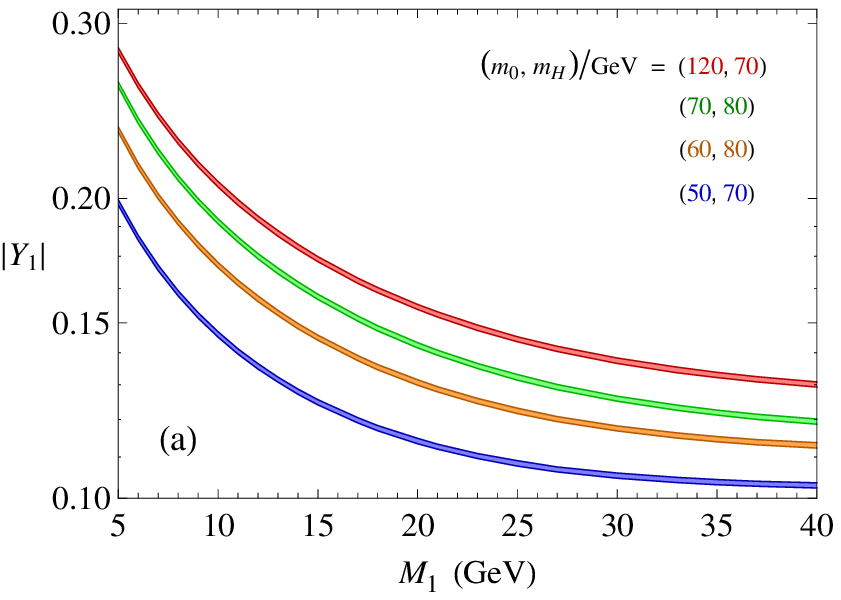} \,
\includegraphics[width=86mm]{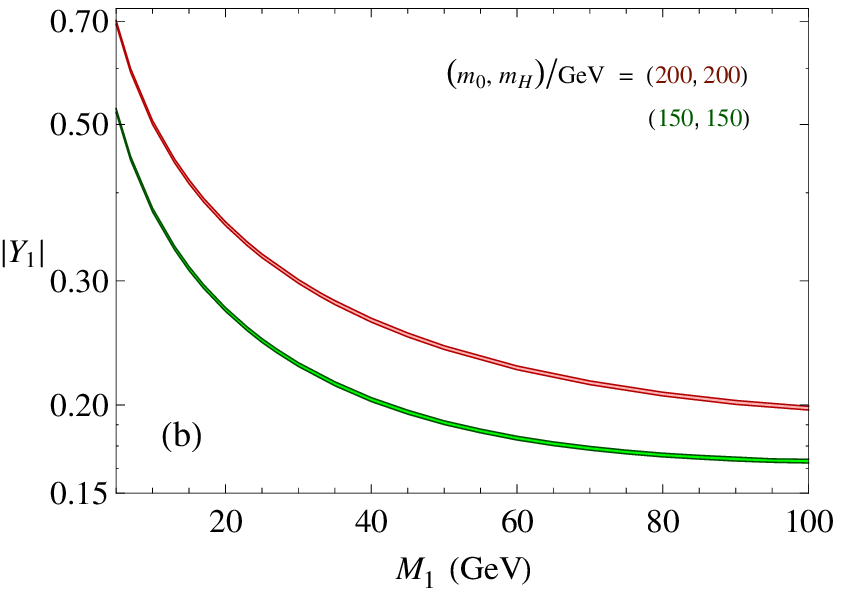} \vspace{-5pt}
\caption{Sample values of Yukawa parameter $|Y_1|$ over two different ranges of $N_1$ mass,
$M_1$, fulfilling the relic density requirement for some selections of the new
neutral and changed scalars' masses $\bigl(m_0^{},m_H^{}\bigr)$.\label{Y1plots}}
\end{figure}

Also important are constraints on \,${\cal Y}_{r1}\propto Y_1$\, from the observed DM
relic abundance, $\Omega$, as we have selected $N_1$ to be the lightest of
the nonstandard particles and play the role of cold~DM.
The dominant contributions to $\Omega$ arise from the $N_1$ annihilations into
\,$\nu_i^{}\nu_j^{}$\, and \,$\ell_i^-\ell_j^+$,\, which are induced at tree level by
$({\cal S,P})$ and $H^\pm$ exchanges, respectively.
Each of them involves diagrams in the $t$ and $u$ channels because of the Majorana
nature of the external neutral fermions.
In Ref.\,\cite{Ho:2013hia} we derived the amplitudes for
\,$N_1^{}N_1^{}\to\nu_i^{}\nu_j^{},\ell_i^-\ell_j^+$\, and computed the corresponding
annihilation rate in order to extract the values of $|Y_1|$ consistent with
the $\Omega$ data supplied by the PDG.
Here we update the allowed ranges of $|Y_1|$ by demanding it to satisfy instead
\,$0.1159\le\Omega\hat h^2\le0.1215$,\, where $\hat h$ is the Hubble constant.
This is the 90\%-confidence-level range of \,$\Omega\hat h^2=0.1187\pm0.0017$\,
which was determined by the Planck Collaboration~\cite{planck}
from the Planck measurement and other data.
We display in Fig.\,\,\ref{Y1plots}(a) some examples of the resulting $|Y_1|$ over
\,$5{\rm\;GeV}\le M_1\le40\;$GeV\, for the solutions in Eq.\,(\ref{123})
and different sets of \,$m_0^{}\simeq m_{\cal S}^{}\simeq m_{\cal P}^{}$\, and $m_H^{}$.
The allowed ranges of $|Y_1|$ in this plot are narrower than those found in
Ref.\,\cite{Ho:2013hia} using the less precise PDG number for~$\Omega$.
In Fig.\,\,\ref{Y1plots}(b) we present examples for a~larger range of $M_1$ which
may be probed at high-energy electron-positron colliders.
More details on the various constraints mentioned only briefly in this section are
available in Ref.\,\cite{Ho:2013hia}.

\section{Scotogenic effects in Higgs decay and $\bm{e^+e^-}$ collisions\label{colliders}}
\vspace{-3pt}

For the $M_1$ range shown in Fig.\,\,\ref{Y1plots}(a), the appropriate values of
\,$M_{2,3}>M_1$,\, and sufficiently low masses of the new scalars, $m_{0,H}^{}$, the Higgs
boson $h$ may decay into final states involving the nonstandard particles.
In Ref.\,\cite{Ho:2013hia}, we considered such decays which proceed from tree-level diagrams,
namely \,$h\to{\cal SS}_{\,}({\cal PP})$\, or
\,$h\to\nu_r^{}N_{k\,}^{}{\cal S}_{\,}({\cal P}),H^\pm\ell_r^\mp N_k^{}$,\, depending on
the daughter particles' masses, over the regions
\,$50{\rm\,GeV}\le m_0^{}\le120$\,GeV\, and \,$70{\rm\,GeV}\le m_H^{}\le120$\,GeV.\,
As we found previously, these exotic decay channels are allowed to have enhanced rates by
the constraints described in the preceding section, including the updated one from the Planck data.
We list several instances of this in the tenth column of Table\,\,\ref{numbers} for
different sets of the mass parameters $m_{0,H}^{}$, $\mu_2^{}$, and $M_{1,2,3}^{}$.
For this table, we have employed the Higgs mass \,$m_h^{}=125.5$\,GeV,\, compatible with
the latest measurements~\cite{atlas:h,cms:h},
and the SM Higgs total width~\,$\Gamma_h^{\rm SM}=4.14\;$MeV\,~\cite{lhctwiki}.
The branching ratio
\,${\cal B}_{\scriptscriptstyle{\cal SP}H}=\Gamma_{\scriptscriptstyle{\cal SP}H}/
\bigl(\Gamma_h^{\rm SM}+\Gamma_{\scriptscriptstyle{\cal SP}H}\bigr)$\,
involves the combined rate $\Gamma_{\scriptscriptstyle{\cal SP}H}$ of all of
the kinematically permitted exotic modes mentioned above~\cite{Ho:2013hia}.
The two numbers on each line under ${\cal B}_{\scriptscriptstyle{\cal SP}H}$
correspond to the two different numbers on the same line in the $\mu_2^{}$ column,
which includes the possibility that $\mu_2^2$ can be negative~\cite{Deshpande:1977rw}.
In the last four rows, \,${\cal B}_{\scriptscriptstyle{\cal SP}H}=0$\, because
these exotic decays of the Higgs cannot happen for the large mass choices.

The last two columns in Table\,\,\ref{numbers} illustrate the impact of the new particles on
the standard decay channels \,$h\to\gamma\gamma$\, and \,$h\to\gamma Z$.\,
These decays are of great interest because they arise from loop diagrams and hence
are sensitive to possible new-physics contributions, which are $H^\pm$ in our case.
Furthermore, these channels are already under investigation at the LHC~\cite{atlas:h,cms:h,h2gz}.
The ratios
\,${\cal R}_{\gamma{\cal V}^{\scriptscriptstyle0}}=
\Gamma(h\to\gamma{\cal V}^{\scriptscriptstyle0})/
\Gamma(h\to\gamma{\cal V}^{\scriptscriptstyle0})_{\rm SM}$\,
for \,${\cal V}^{\scriptscriptstyle0}=\gamma,Z$\,
would thus signal new physics if they are unambiguously measured to deviate from unity.

\begin{table}[b]
\caption{Sample values of mass parameters $m_{0,H}^{}$, $\mu_2^{}$, and $M_{1,2,3}$,
and Yukawa constants $Y_{1,2,3}$ satisfying the constraints discussed in
Section~\ref{constraints} and the resulting branching ratios
${\cal B}_{\scriptscriptstyle{\cal SP}H}$ of the Higgs decay into final states containing
$\cal S, P$, or $H^\pm$ and ratios ${\cal R}_{\gamma{\cal V}^0}$ of
$\Gamma(h\to\gamma{\cal V}^{\scriptscriptstyle0})$ to its SM value for
\,${\cal V}^{\scriptscriptstyle0}=\gamma,Z$.\label{numbers}\medskip}
\begin{tabular}{|ccccccccc|ccc|} \hline
$~\frac{\displaystyle m_0^{}}{\scriptstyle\rm GeV_{\vphantom{\int_|}}}~$ &
$\frac{\displaystyle m_H^{}}{\scriptstyle\rm GeV}$ &
$\frac{\displaystyle\mu_2^{}}{\scriptstyle\rm GeV}$ & \,
$\frac{\displaystyle M_1^{}}{\scriptstyle\rm GeV}$ \, &
$\frac{\displaystyle M_2^{}}{\scriptstyle\rm GeV}$ & \,
$\frac{\displaystyle M_3^{}}{\scriptstyle\rm GeV}$ \, & $Y_1^{}$ & $Y_2^{}$ & $Y_3^{}$ &
\raisebox{-4pt}{\small$\stackrel{\displaystyle{\cal B}_{\scriptscriptstyle{\cal SP}H}
{}^{\vphantom{0}}}{\scriptstyle(\%)^{\vphantom{\int^|}}}$}
& ${\cal R}_{\gamma\gamma}^{}$ & ${\cal R}_{\gamma Z}^{}$ \\
\hline\hline
 50 &  70 &  46 (47)   &   9 &  14 &  64 & \, 0.152 \, & 0.363 & \, 0.642 \, & \,
$\vphantom{\int^|}$20 (14) \, & \, 0.89 (0.89) \, & 0.95 (0.95)$\vphantom{\int^{|^|}}$
\\
 60 &  80 &  54 (56)   &  10 &  15 &  72 & 0.171 & 0.410 & 0.703 &
$\vphantom{\int^|}$26 (14) &  0.91 (0.92) & 0.96 (0.97)
\\
 70 &  80 & 113 $(7i)$  & 12 & 18 & 79 & 0.175 & 0.422 & 0.740 &
$\vphantom{\int^|}$24 (12) & 1.2 (0.84) & 1.1 (0.93)
\\
120 &  70 & 123 (111)   & 20 & 29 & 85 & 0.155 & 0.380 & 0.712 &
$\vphantom{\int_|^|}$20 (12) & 1.5 (1.3) & 1.2 (1.1) \\
\hline
 50 &  85 & 54 (53)      & 35 & 51 & 143 & 0.107 & 0.262 & 0.603 &
$\vphantom{\int^{|^|}}$21 (13) & 0.91 (0.91) & \,0.96 (0.96)\,
\\
 50 &  90 & 46 (47)      & 30 & 43 & 125 & 0.110 & 0.264 & 0.575 &
$\vphantom{\int^|}$18 (11) & 0.90 (0.90) & 0.96 (0.96)
\\
 65 &  90 & 140  $(70i)$ & 40 & 57 & 153 & 0.119 & 0.293 & 0.658 &
$\vphantom{\int^|}$25 (11) & 1.2 (0.79) & 1.1 (0.91)
\\
 70 &  85 & \,199 $(135i)$\, & 50 &  71 & 178 & 0.119 & 0.300 & 0.707 &
$\vphantom{\int_|^|}$\,0.6 (0.3)\, & 1.8 (0.54) & 1.3 (0.80)
\\ \hline
150 & 150 & 80 $(280)$ &  50 &  72 & 181 & 0.188 & 0.452 & 0.917 &
$\vphantom{\int^{|^|}}$0 (0) & 0.92 (1.3) & 0.97 (1.1)
\\
150 & 150 & 90 $(290)$ & 100 & 142 & 277 & 0.167 & 0.415 & 0.947 &
$\vphantom{\int^|}$0 (0) & 0.93 (1.3) & 0.97 (1.1)
\\
200 & 200 & 80 $(330)$ &  50 &  75 & 220 & 0.241 & 0.578 & 1.131 &
$\vphantom{\int^|}$0 (0) & 0.91 (1.2) & 0.97 (1.1)
\\
200 & 200 & 70 $(340)$ & 100 & 143 & 265 & 0.199 & 0.477 & 1.027 &
$\vphantom{\int_|^|}$0 (0) & 0.91 (1.2) & 0.97 (1.1)
\\ \hline \end{tabular}
\end{table}

The predictions for ${\cal B}_{\scriptscriptstyle{\cal SP}H}$ and ${\cal R}_{\gamma\gamma}$
in Table\,\,\ref{numbers} can already be tested experimentally.
Recent analyses~\cite{h2inv} have determined that the present Higgs data allow
the branching ratio of its nonstandard decays into invisible or undetected final states
to reach 22\% at the 95\% confidence level if the Higgs production mechanism is~SM-like,
which is the case in the scotogenic model.
This restriction is not yet severe for ${\cal B}_{\scriptscriptstyle{\cal SP}H}$ and can
be readily avoided by changing~$\mu_2^{}$, as can be viewed in the table.
For~\,$h\to\gamma\gamma$,\, which has been detected, unlike the $\gamma Z$
channel~\cite{h2gz}, the prediction can be compared to observation.
The measurements of the signal strength for \,$h\to\gamma\gamma$\, by the ATLAS and CMS
Collaborations are~\,$\sigma/\sigma_{\rm SM}^{}=1.55_{-0.28}^{+0.33}$\,~\cite{atlas:h}
and \,$\sigma/\sigma_{\rm SM}^{}=0.77\pm0.27$\,~\cite{cms:h}, respectively.
Evidently, the majority of the ${\cal R}_{\gamma\gamma}$ numbers are
in agreement with one or the other of these LHC results.
Pending an experimental consensus on this decay mode and the advent of complementary
information from the future detection of \,$h\to\gamma Z$,\, we are motivated to pursue
other means to probe the model to a~greater degree.

The new sector of the model being leptophilic, one may want to look into extra tests on
it by means of electron-positron scattering.
Below we demonstrate that potentially significant restraints on the model are
indeed available from past measurements at LEP\,II.
Since the ILC may become a~reality in the not-too-distant future, providing $e^+e^-$
scattering experiments at higher energies and with better precision,
we also make some estimates and comments relevant to it.
In the rest of this section, we focus on scotogenic contributions to $e^+e^-$ collisions
into a~pair of charged leptons plus missing energy and into a photon plus missing energy.

\subsection{$\bm{e^+e^-\to H^+H^-\to\ell^+\ell^{\prime-\;}\slash\!\!\!\!\!E}$\label{ee2HH2llE}}

The amplitude for \,$e^+(p_+)_{\,}e^-(p_-)\to H^+(q_+)_{\,}H^-(q_-)$,\, which comes from
$\gamma$- and $Z$-mediated diagrams in the $s$ channel and $N_k$-mediated diagrams
in the $t$ channel, follows from Eqs.\,\,(\ref{Lphieta}) and~(\ref{LN}).
It can be expressed as
\begin{eqnarray} \label{Mee2HH}
{\cal M}_{e\bar e\to H\bar H}^{} \,\,=\,\,
\frac{-2e^2\,\bar v_{e^+}^{}\!\not{\!q}_-^{}u_{e^-}^{}}{s}
\,-\,
\frac{2\,\bar v_{e^+}^{}\!\not{\!q}_-^{}\bigl(g_L^2P_L^{}+g_L^{}g_R^{}P_R^{}\bigr)
u_{e^-}^{}}{s-m_Z^2+i\Gamma_Z^{}m_Z^{}}
\,+\,
\raisebox{0.5ex}{\footnotesize$\displaystyle\sum_k$}\,
\frac{|{\cal Y}_{1k}|^2\,\bar v_{e^+}^{}\!\not{\!q}_-^{}P_L^{}u_{e^-}^{}}{M_k^2-t} ~, ~~
\end{eqnarray}
where $u_e^{}$ and $v_e^{}$ are Dirac spinors, \,$s=(p_++p_-)^2$,\, \,$t=(p_+-q_+)^2$,\,
\,$g_L^{}=g\bigl(s_{\rm w}^2-1/2\bigr)/c_{\rm w}^{}$,\,
\,$g_R^{}=g_{\,}s_{\rm w}^2/c_{\rm w}^{}$,\, and \,$P_L^{}=\frac{1}{2}(1-\gamma_5)$.\,
We have relegated the resulting cross-section, $\sigma_{e\bar e\to H\bar H}$,
to Eq.\,(\ref{sigma_ee2HH}) in~Appendix\,\,\ref{csr}.

After their production, $H^\pm$ will decay into \,$\ell_o^\pm N_k^{}$\, if
\,$m_H^{}>m_{\ell_o}^{}+M_k^{}$.\,
For \,$k=2$ or 3,\, the decays \,$N_k^{}\to\ell_{r\,}^\pm H^\mp$\, and
\,$N_k^{}\to\nu{\cal S},\nu\cal P$\, may occur,
followed, respectively, by \,$H^\pm\to\ell_s^\pm N_l^{}$\, if \,$m_H^{}>m_{\ell_s}+M_l$\, and
\,${\cal S,P}\to\nu N_l^{}$.\,
If these two-body channels of $N_k$ are not open, it will instead undergo
\,$N_k^{}\to\nu_o^{}\nu_r^{}N_l^{}$\, and possibly \,$N_k^{}\to\ell_o^-\ell_r^+N_l^{}$.\,
We have collected the expressions for the rates of these various decays of $H^\pm$, $\cal S$,
$\cal P$, and $N_k$ in Appendix\,\,\ref{csr}.\footnote{In this paper we do not consider
scenarios with \,$|m_H^{}-m_{\cal S,P}^{}|\ge m_{W,Z}^{}$,\, in which the new scalars may also
be detectable through other two-body decays, like \,$H^\pm\to{\cal S(P)}W^\pm$.\,
Such a possibility has been discussed in the context of the inert doublet
model~\cite{idm,Aoki:2013lhm} without $N_k$.}
In the final states of the decays just mentioned, $N_l$ will no longer decay if \,$l=1$.\,

Thus, since $N_1$ is DM, the channel
\,$e^+e^-\to\ell^+\ell^{\prime-\;}\slash\!\!\!\!E$\, with missing energy \,$\slash\!\!\!\!E$\,
in the final state receives the scotogenic contribution
\,$e^+e^-\to H^+H^-\to\ell^+\ell^{\prime-\;}\slash\!\!\!\!E$.\,
We can write its cross section as
\begin{eqnarray} \label{ee2HH2llNN}
\sigma_{e\bar e\to H\bar H\to\ell\bar\ell'\slash\!\!\!\!E}^{} \,\,=\,\,
\sigma_{e\bar e\to H\bar H}^{}\,
\Bigl(\raisebox{2pt}{\scriptsize$\displaystyle\sum_{\mbox{\scriptsize$r$}}$}\,
{\cal B}(H\to\ell_r\;\slash\!\!\!\!E)\Bigr)\raisebox{4pt}{$^2$}
\end{eqnarray}
with the branching ratios
\begin{eqnarray} & \displaystyle \label{H2lE}
{\cal B}(H\to\ell_r\;\slash\!\!\!\!E) \,\,=\,\, {\cal B}(H\to\ell_r N_1)
+ {\cal B}(H\to\ell_r N_2)\,{\cal B}_{21} + {\cal B}(H\to\ell_r N_3)\,{\cal B}_{31} ~,
& ~ ~ ~ \nonumber \\ \label{Bij} & \displaystyle
{\cal B}_{21} \,\,=\,\, {\cal B}(N_2\to\nu\nu'N_1) ~, \hspace{5ex}
{\cal B}_{31} \,\,=\,\, {\cal B}(N_3\to\nu\nu'N_1)
+ {\cal B}(N_3\to\nu\nu'N_2)\,{\cal B}_{21} \vphantom{\big|^{\int}} ~,
\end{eqnarray}
where
\,${\cal B}(N_k\to\nu\nu'N_l)=\mbox{\small$\sum_{\scriptscriptstyle\hat\eta=\cal S,P\,}$}
\bigl(\Gamma_{N_k\to\nu\hat\eta}/\Gamma_{N_k}\bigr)\Gamma_{\hat\eta\to\nu N_l}/\Gamma_{\hat\eta}$\,
or \,$\Gamma_{N_k\to\nu\nu'N_l}/\Gamma_{N_k}$\, depending on the masses.
Any of the terms in ${\cal B}(H\to\ell_r\;\slash\!\!\!\!E)$ would be absent if
kinematically forbidden.

Since according to Eq.\,(\ref{Lphieta}) the $Z$ boson can couple to $\cal S$ and $\cal P$,
it can mediate \,$e^+e^-\to\cal SP$\, in~the~$s$~channel.
This transition is experimentally unobservable if $\cal S$ and $\cal P$ each decay
(sequentially) into an $N_1$ along with one or more $\nu$s, as all of these fermions are
invisible.\footnote{Without observable events, due to the absence of detectable
particles in the final state, an empirical cross-section would not
be available to check the theory~\cite{Ma:1978yf,Konar:2009ae}.\smallskip}
On the other~hand,~if only one member of the $\cal SP$ pair undergoes such a decay, while the other
member decays into \,$\ell^+\ell^{\prime-}N_1$\, and one or more $\nu$s, then
\,$e^+e^-\to\cal SP$\, will also contribute to the 
\,\mbox{$\ell^+\ell^{\prime-\,}\slash\!\!\!\!E$\, final state. With} 
\,$m_{\cal S}^{}\simeq m_{\cal P}^{}\simeq m_0^{}$,\, we can write the cross section of this
contribution as
\,$\sigma_{e\bar e\to{\cal SP}\to\ell\bar\ell'\,\slash\!\!\!\!E}=2_{\,}
\sigma_{e\bar e\to\cal SP}\, {\cal B}(\hat\eta\to\slash\!\!\!\!E)\,
{\cal B}\bigl(\hat\eta\to\ell\bar\ell'\slash\!\!\!\!E\bigr)$,\,
where $\sigma_{e\bar e\to\cal SP}$ is given in~Appendix\,\,\ref{csr} for completeness,
\,$\hat\eta=\cal S$ or $\cal P$,\,
\,${\cal B}(\hat\eta\to\slash\!\!\!\!E)=
{\cal B}(\hat\eta\to\nu N_1) + {\cal B}(\hat\eta\to\nu N_2)\,{\cal B}_{21}
+ {\cal B}(\hat\eta\to\nu N_3)\,{\cal B}_{31}$,\,
and~\,${\cal B}\bigl(\hat\eta\to\ell\bar\ell'\slash\!\!\!\!E\bigr)=
{\cal B}(\hat\eta\to\nu N_2)\, {\cal B}\bigl(N_2\to\ell\bar\ell'\slash{\!\!\!\!E}\bigr)
+ {\cal B}(\hat\eta\to\nu N_3)\, {\cal B}\bigl(N_3\to\ell\bar\ell'\slash{\!\!\!\!E}\bigr)$.\,
Having more powers of the branching ratios,
$\sigma_{e\bar e\to{\cal SP}\to\ell\bar\ell'\,\slash\!\!\!\!E\,}$ can be expected to be
suppressed with respect to~$\sigma_{e\bar e\to H\bar H\to\ell\bar\ell'\,\slash\!\!\!\!E}$.
This turns out to be the case for the parameter choices in our illustrations, the suppression
factors being a few or more.
The impact of $\sigma_{e\bar e\to{\cal SP}\to\ell\bar\ell'\,\slash\!\!\!\!E\,}$ on
$\sigma_{e\bar e\to H\bar H\to\ell\bar\ell'\,\slash\!\!\!\!E}$ is actually further subdued
because the angular distributions of the final lepton pairs, $\ell^+\ell^{\prime-}$, in
the two processes are generally very different.
For these reasons, hereafter we neglect the effect of
\,$e^+e^-\to{\cal SP}\to\ell^+\ell^{\prime-\,}\slash\!\!\!\!E$\, in examining
\,$e^+e^-\to H^+H^-\to\ell^+\ell^{\prime-\,}\slash\!\!\!\!E$.\,

The process \,$e^+e^-\to\ell^+\ell^{\prime-\,}\slash\!\!\!\!E$\, with the final charged leptons
not originating from the same particle has been well measured at LEP\,II~\cite{lep:ee2llnn}.
The experimental values of its cross section at center-of-mass (c.m.) energies
\,$\sqrt s\simeq183$-208$\;$GeV\, vary from about 1.4 to 2.5~pb with errors ranging
mostly between 10\% and 20\%.
Except for several of them, the measurements are consistent at the one-sigma level with
the SM prediction for \,$e^+e^-\to W^+W^-\to\nu\nu'\ell^+\ell^{\prime-}$,\,
summed over all of the final leptons.
Accordingly, we may demand that
\,$\sigma_{e\bar e\to H\bar H\to\ell\bar\ell'\,\slash\!\!\!\!E}<0.3$\;pb.\,

To get some indications as to which of the examples in Table\,\,\ref{numbers} can meet this
condition, we present the cross sections in Table\,\,\ref{csee2llE} at the c.m.
energies \,$\sqrt s=183,196,207$~GeV\, representing the LEP\,II range.
Obviously the parameter values yielding the cross sections at these energies
in the first four rows are disfavored by the LEP\,II data.
In contrast, the corresponding numbers in the second four rows can fulfill the imposed bound,
due to the relatively larger $m_H^{}$ and $M_k$ and smaller~$Y_k$.
Interestingly, in these latter examples, the Higgs exotic decays into the scotogenic
particles can mostly still happen with nonnegligible rates, as illustrated by
their ${\cal B}_{\scriptscriptstyle{\cal SP}H}$ entries in the second four rows
of Table\,\,\ref{numbers}.

\begin{table}[t] \vspace{-2ex}
\caption{Cross section of \,$e^+e^-\to H^+H^-\to\ell^+\ell^{\prime-\;}\slash\!\!\!\!E$\,
corresponding to the examples in Table\,\,\ref{numbers} for c.m. energies
\,$\sqrt s=183,196,207,250,500,1000$~GeV,\, with both $H^\pm$ being
on-shell.\label{csee2llE}\medskip} 
\begin{tabular}{|cccccccc|ccc|ccc|} \hline
\multirow{2}{*}{$~\frac{\displaystyle m_0^{}}{\rm GeV}~$} &
\multirow{2}{*}{$\frac{\displaystyle m_H^{}}{\rm GeV}$} &
\multirow{2}{*}{\, $\frac{\displaystyle M_1^{}}{\rm GeV}$} \, &
\multirow{2}{*}{$\frac{\displaystyle M_2^{}}{\rm GeV}$} & \,
\multirow{2}{*}{$\frac{\displaystyle M_3^{}}{\rm GeV}$} \, &
\multirow{2}{*}{$Y_1^{}$} & \multirow{2}{*}{$Y_2^{}$} & \multirow{2}{*}{$Y_3^{}$} &
\multicolumn{6}{c|}{$\sigma_{e\bar e\to H\bar H\to\ell\bar\ell'_{\,}\slash\!\!\!\!E}
\vphantom{|_{\int}^|}$ (pb)} \\
\cline{9-14}
& & & & & & & & \scriptsize 183 & \scriptsize 196 & \scriptsize 207 &
\scriptsize 250 & \scriptsize 500 & \scriptsize 1000 \\
\hline\hline
 50 &  70 &   9 &  14 &  64 & \, 0.152 \, & 0.363 & \, 0.642 \, &
$\vphantom{\int^{|^|}}$4.0 & 4.8 & 5.3 & 6.2 & 3.8 & 1.4
\\
 60 &  80 &  10 &  15 &  72 & 0.171 & 0.410 & 0.703 &
$\vphantom{\int^|}$2.5 & 3.9 & 5.0 & 7.2 & 5.5 & 2.1
\\
 70 &  80 &  12 & 18 & 79 & 0.175 & 0.422 & 0.740 &
$\vphantom{\int^|}$2.3 & 3.7 & 4.7 & 6.8 & 5.0 & 1.9
\\
120 &  70 &  20 & 29 & 85 & 0.155 & 0.380 & 0.712 &
$\vphantom{\int_|^|}$1.4 & 1.7 & 1.9 & 2.2 & 1.3 & 0.49
\\ \hline
 50 &  85 &  35 & 51 & 143 & 0.107 & 0.262 & 0.603 &
$\vphantom{\int^{|^|}}$\, 0.06 \, & 0.16 & \, 0.24 \, & \, 0.49 \, & 0.52 & \, 0.24 \,
\\
 50 &  90 &  30 & 43 & 125 & 0.110 & 0.264 & 0.575 &
$\vphantom{\int^|}$0.01 & 0.10 & 0.19 & 0.51 & 0.61 & 0.27
\\
 65 &  90 &  40 & 57 & 153 & 0.119 & 0.293 & 0.658 &
$\vphantom{\int^|}$0.01 & 0.12 & 0.23 & 0.59 & 0.71 & 0.33
\\
 70 &  85 &  50 & 71 & 178 & 0.119 & 0.300 & 0.707 &
$\vphantom{\int_|^|}$0.08 & 0.19 & 0.29 & 0.57 & 0.66 & 0.32
\\ \hline
150 & 150 &  50 &  72 & 181 & 0.188 & 0.452 & 0.917 &
$\vphantom{\int^{|^|}}$  0  & 0    &   0  &  0   &  1.8 &  1.2
\\
150 & 150 & 100 & 142 & 277 & 0.167 & 0.415 & 0.947 &
$\vphantom{\int^|}$  0  & 0    &   0  &  0   & 0.96 & 0.76
\\
200 & 200 &  50 &  75 & 220 & 0.241 & 0.578 & 1.131 &
$\vphantom{\int^|}$  0  & 0    &   0  &  0   &  2.0 &  2.8
\\
200 & 200 & 100 & 143 & 265 & 0.199 & 0.477 & 1.027 &
$\vphantom{\int_|^|}$  0  & 0    &   0  &  0   & 0.69 &  1.1
\\ \hline \end{tabular}
\end{table}

For the parameter space that can evade the LEP\,II restrictions and has room for
the scotogenic decays of the Higgs compatible with LHC data, further tests are
potentially available at a future higher-energy $e^+e^-$ collider, such as the ILC~\cite{ilc}.
Moreover, $H^\pm$ which are too heavy to have been produced at LEP\,II may be within
the reach of the ILC.\footnote{\baselineskip=13pt%
The potential reach of the ILC to measure \,$e^+e^-\to H^+H^-$\,
in the inert doublet model~\cite{Deshpande:1977rw,idm} without $N_k$, or in the scotogenic model
with $\cal S$ being the DM candidate and $N_k$ very heavy, has recently been studied
in~Ref.\,\cite{Aoki:2013lhm}.}
The last three columns of Table\,\,\ref{csee2llE} show a number of predictions for
$\sigma_{e\bar e\to H\bar H\to\ell\bar\ell'\,\slash\!\!\!\!E\,}$
at some of the proposed ILC energies.
The predictions are to be compared with the SM cross-sections
\begin{eqnarray}
\sigma_{e\bar e\to W\bar W\to\nu\nu'\ell\bar\ell'} \,\,=\,\, 1.7,\,0.8,\,0.3~\rm pb
\end{eqnarray}
which are the tree-level values at \,$\sqrt s=250,500,1000$~GeV,\, respectively.
Since the scotogenic contributions are of roughly similar order to, or substantially exceed, the SM
ones, we can conclude that experiments on \,$e^+e^-\to\ell^+\ell^{\prime-\;}\slash\!\!\!\!E$\,
at the ILC have the potential to discover scotogenic signals or impose stringent
limits on the parameter regions examined in this paper.

Once $H^\pm$ are discovered, precise measurements on their decay modes,
especially \,$H^\pm\to\ell_r^\pm N_k^{}$,\, will help uncover the flavor structure of
the Yukawa interactions of the new particles.
Specifically, as Eq.\,(\ref{H2lN}) indicates, ratios of the magnitudes of Yukawa couplings
${\cal Y}_{rk}$ can be inferred from the ratios of the experimental branching ratios of these
two-body decays.
At $e^+e^-$ colliders, such ratios can be measured after sufficient data are accumulated to
allow the identification of the lepton flavors in
the \,$\ell^+\ell^{\prime-\;}\slash\!\!\!\!E$\, signal events.
However, it may be difficult to extract clearly the individual $|{\cal Y}_{rk}|$ themselves
because \,$e^+e^-\to H^+H^-$\, is induced not only by $\gamma$- and $Z$-mediated diagrams,
but also by $N_k$-mediated diagrams which involve~${\cal Y}_{1k}$.
On the other hand, at the LHC both the relative and absolute values of $|{\cal Y}_{rk}|$
are measurable if enough statistics are available, as the main production channel is the quark
annihilation \,$q\bar q\to H^+H^-$\, via $\gamma$ and $Z$ exchanges only.
The acquired data on \,$H^\pm\to\ell_r^\pm N_k^{}$\, will, in addition, reveal the masses of $N_k$.
All of this information on ${\cal Y}_{rk}$ and $M_k$, plus the masses of the new scalars,
is crucial because they also determine the light neutrinos' mass matrix and the rates of
the flavor-changing decays~\,$\ell_o\to\ell_r\gamma$,\, as well as the relic density of
the DM particle~$N_1$.
In other words, a~good amount of experimental data on the various quantities which are
functions of ${\cal Y}_{rk}$ and the new particles' masses will serve to check the predictions,
and hence the self-consistency, of the model.\footnote{\baselineskip=13pt%
Much of the discussion in this paragraph
also applies to some other scenarios of one-loop radiative neutrino mass in which the neutrino
and DM sectors are intimately connected, such as the model proposed in~Ref.\,\cite{slim}.
Its LHC phenomenology is the focus of Ref.\,\cite{Farzan:2010fw}, which provides a detailed
analysis on the possibility of searching for a~signal in
\,$\ell^+\ell^{\prime-}\slash\!\!\!\!E$\, final-states.\vspace{-3ex}}

\subsection{$\bm{e^+e^-\to\gamma N N',\gamma{\cal SP}\to\gamma\;\slash\!\!\!\!\!E}$\label{ee2HH}}

Another kind of scotogenic effect that may be observable at $e^+e^-$ colliders is
\,$e^+e^-\to\gamma N_j N_k$,\, which, if \,$j,k>1$,\, is followed by $N_{j,k}$ decaying
(sequentially) into $N_1$ plus light neutrinos.
This is generated by $H$-exchange diagrams with the photon radiated off the $e^\pm$ lines.
We have written down the scattering amplitude, which depends on ${\cal Y}_{1j,1k}$, and sketched
the calculation of the cross section, $\sigma_{e\bar e\to \gamma N_j N_k}$, in Appendix\,\,\ref{csr}.
In view of the Majorana nature of $N_{j,k}$, we can express its contribution to
the monophoton production process \,$e^+e^-\to\gamma\,\slash\!\!\!\!E$\, as
\begin{eqnarray} \displaystyle \label{cs_ee2gNN}
\sigma_{e\bar e\to \gamma N N'\to\gamma\,\slash\!\!\!\!E}^{} \,\,=\,\,
\raisebox{2pt}{\footnotesize$\displaystyle\sum_{
\mbox{$\stackrel{j,\,k\,=1}{\scriptstyle j{\scriptscriptstyle\,\le\,} k}$}}^3$}\,
\sigma_{e\bar e\to \gamma N_j N_k}^{}\,{\cal B}_{j1}^{}\,{\cal B}_{k1}^{} ~,
\end{eqnarray}
where ${\cal B}_{21,31}$ are defined in Eq.\,(\ref{Bij}) and \,${\cal B}_{11}=1$.\,
Any of the terms in this sum would vanish if kinematically forbidden.

There is an additional scotogenic contribution to
\,$e^+e^-\to\gamma\,\slash\!\!\!\!E$,\, namely \,$e^+e^-\to\gamma{\cal SP}$\,
induced by $Z$-mediated diagrams with the photon being emitted from the $e^\pm$ legs.
We have outlined the computation of its cross section, $\sigma_{e\bar e\to\gamma\cal SP}$,
in Appendix\,\,\ref{csr}.
The $\gamma\,\slash\!\!\!\!E$ final-state is reached when $\cal S$ and
$\cal P$ each decay (sequentially) into $N_1$ and one or more light neutrinos.
Putting things together, we arrive at the cross section
\begin{eqnarray} \label{cs_ee2gSP}
\sigma_{e\bar e\to\gamma{\cal SP}\to\gamma\,\slash\!\!\!\!E}^{} \,\,=\,\,
\sigma_{e\bar e\to\gamma\cal SP}^{}\,\bigl({\cal B}(\hat\eta\to\slash\!\!\!\!E)\bigr)^2
\end{eqnarray}
with the branching ratio
\,${\cal B}(\hat\eta\to\slash\!\!\!\!E)={\cal B}(\hat\eta\to\nu N_1)
+ {\cal B}(\hat\eta\to\nu N_2)\,{\cal B}_{21}+{\cal B}(\hat\eta\to\nu N_3)\,{\cal B}_{31}$.\,
As it turns out, $\sigma_{e\bar e\to\gamma{\cal SP}\to\gamma\,\slash\!\!\!\!E\,}$ is numerically
less important than $\sigma_{e\bar e\to \gamma N N'\to\gamma\,\slash\!\!\!\!E\,}$ for the mass and
coupling values in our examples.

Much experimental work on \,$e^+e^-\to\gamma\,\slash\!\!\!\!E$\, has also been performed at
LEP\,II to study the neutrino counting reaction
\,$e^+e^-\to\gamma\nu\bar\nu$\, in the SM and also to search for long-lived or stable
new particles~\cite{lep:ee2gnn}.
The measured cross-sections at \,$\sqrt s\simeq130$-207~GeV,\, with errors mainly between
5\% and 20\%, vary not only with $\sqrt s$, but also with the experimental cuts on
the photon energy $E_\gamma$ and angle $\theta_\gamma$ relative to the beam direction.
From a collection of these data~\cite{lep:ee2gnn} tabulated in~Ref.\,\cite{Chiang:2012ww},
one can see that the experimental and SM values of the cross section agree with each other
at the one-sigma level, except for several of them.

Comparing with the LEP\,II results on \,$e^+e^-\to\gamma\,\slash\!\!\!\!E$,\, we find
that for the parameter ranges that escape the bounds from
\,$e^+e^-\to\ell^+\ell^{\prime-\,}\slash\!\!\!\!E$\, data discussed in the previous subsection
the scotogenic contributions to \,$e^+e^-\to\gamma\,\slash\!\!\!\!E$\, at LEP\,II energies
do not yield significant effects.
The scotogenic contributions are even small compared to the experimental errors.
Consequently, we need to turn to the ILC in order to explore the possibility of seeing
the desired signals.\footnote{Similar situations may arise in some other radiative neutrino
mass models with fermionic DM~\cite{Ahriche:2013zwa,Law:2013saa} and more generally in models
with nonnegligible effective DM-electron couplings~\cite{Konar:2009ae}.}

After appropriate cuts on the photon energy and angle are imposed, the main background is
\,$e^+e^-\to\gamma\nu\bar\nu$\, in the SM which can be calculated with formulas
available in the literature~\cite{Chiang:2012ww,sm:ee2gnn}.
Among the examples in Table\,\,\ref{csee2llE}, we obtain a few that produce contributions to
\,$e^+e^-\to\gamma\,\slash\!\!\!\!E$\, which are not negligible compared to the background
at ILC energies.
We display the results in Table\,\,\ref{csee2gE} where the cuts used are specified.
The entries for \,$\sqrt s=250,500,1000$~GeV\, are to be compared to the SM numbers in
the bottom row.
Although the cross sections of
\,\mbox{$e^+e^-\to\gamma N N',\gamma{\cal SP}\to\gamma\,\slash\!\!\!\!E$}\,
in the first two rows are below 4\% of the background, the ones in the next two
rows can reach about 7\% to 13\%, notably at \,$\sqrt s=500,1000$~GeV.\,
Assuming that the proposed integrated luminosities of \,500 and 1000~fb$^{-1}$\, at these
energies~\cite{ilc}, respectively, are achievable, we may expect that there will be enough
events to distinguish signals from backgrounds.
If that is the case, then these examples have illustrated that the information to be gained
from the ILC data on \,\mbox{$e^+e^-\to\gamma\,\slash\!\!\!\!E$}\, is complementary to that from
\,$e^+e^-\to\ell^+\ell^{\prime-\,}\slash\!\!\!\!E$\, in probing the scotogenic model further.
Especially, if a new-physics hint is detected in the \,$\ell^+\ell^{\prime-\,}\slash\!\!\!\!E$\,
events, the $\gamma\,\slash\!\!\!\!E$ measurement could serve to offer some cross-checks,
but the observation of a nonstandard signal in only the $\gamma\,\slash\!\!\!\!E$ data would
likely disfavor the scenario discussed above.

\begin{table}[h]
\caption{Cross section of \,$e^+e^-\to\gamma N N',\gamma{\cal SP}\to\gamma\,\slash\!\!\!\!E$,\,
in fb, for the parameter values in some of the examples in Table\,\,\ref{csee2llE} at
c.m. energies \,$\sqrt s=250,500,1000$~GeV.\,
The two terms in each of the sums correspond to the $\gamma N N'$ and $\gamma{\cal SP}$
contributions, respectively.
The cuts applied to the photon energy and angle relative to the incident electron's direction are
\,$E_\gamma\sin\theta_\gamma\ge0.15\,\sqrt s$\, and \,$|\!\cos\theta_\gamma|\le0.7$,\, as well as
\,$E_\gamma\le\min\bigl(0.45\,\sqrt s,\,E_\gamma^{\rm max}\bigr)$,\, where $E_\gamma^{\rm max}$
is related to the $N_{k,l\,}$ [$\cal S,P$] masses by~Eq.\,(\ref{Egrange})~[Eq.\,(\ref{egrange})].
In the bottom row are the corresponding numbers for \,$e^+e^-\to\gamma\nu\bar\nu$\, in
the SM, with the same cuts except~\,$E_\gamma\le0.45\,\sqrt s$.\label{csee2gE}\medskip}
\begin{tabular}{|ccc ccc cc|ccc|} \hline
\multirow{2}{*}{$~\frac{\displaystyle m_0^{}}{\rm GeV}~$} &
\multirow{2}{*}{$\frac{\displaystyle m_H^{}}{\rm GeV}$} &
\multirow{2}{*}{\, $\frac{\displaystyle M_1^{}}{\rm GeV}$} \, &
\multirow{2}{*}{$\frac{\displaystyle M_2^{}}{\rm GeV}$} & \,
\multirow{2}{*}{$\frac{\displaystyle M_3^{}}{\rm GeV}$} \, &
\multirow{2}{*}{$Y_1^{}$} & \multirow{2}{*}{$Y_2^{}$} & \multirow{2}{*}{$Y_3^{}$} &
\multicolumn{3}{c|}{$\sigma_{e\bar e\to\gamma N N',\gamma{\cal SP}\to
\gamma\,\slash\!\!\!\!E}\vphantom{|_{\int}^|}$ (fb)} \\
\cline{9-11}
& & & & & & & & \scriptsize 250 ~ \, & \scriptsize 500 ~ & \scriptsize 1000 ~ \\
\hline\hline
 65 &  90 &  40 & 57 & 153 & 0.119 & 0.293 & 0.658 &
$\vphantom{\int^{|^|}}3.3+0.76$ & $2.5+0.48$ & $1.0+0.14$
\\
 70 &  85 &  50 & 71 & 178 & 0.119 & 0.300 & 0.707 &
$\vphantom{\int^|}2.7+0.54$ & $2.7+0.46$ & $1.1+0.14$
\\
150 & 150 &  50 &  72 & 181 & 0.188 & 0.452 & 0.917 &
$\vphantom{\int^|}4.7+0$ ~ \, & $6.7+0.04$ & $3.5+0.06$
\\
200 & 200 &  50 &  75 & 220 & \, 0.241 \, & 0.578 & \, 1.131 \, & ~
$\vphantom{\int_|^|}6.3+0$ ~ ~ \, & \,$12.2+0.00$ \, & ~ $7.6+0.03$ ~
\\ \hline \hline
\multicolumn{8}{|c|}
{$\sigma_{e\bar e\to\gamma\nu\bar\nu}^{\textsc{sm}}\vphantom{|_{\int}^{\int}}$ (fb)}
& 613 \, & \, 95.5 \, & \, 61.1 \,
\\ \hline
\end{tabular}
\end{table}

\section{Conclusions\label{summary}}

We have investigated the possibility of employing $e^+e^-$ colliders to
provide additional tests on the scotogenic model of radiative neutrino mass.
This study continues our previous work which addressed the Higgs boson undergoing
exotic decays into the nonstandard particles of the same model.
Unlike before, here we adopt the PDG parametrization of the neutrino-mixing matrix
and derive exact solutions for the Yukawa couplings of the new particles in
terms of three free (complex) parameters.
Accordingly, the Yukawa results are consistent with the measured elements of
the mixing matrix.
We select one set of such solutions to be used in our numerical computation.
As before, we assume that the lightest one of the new fermions is the cold DM candidate.
Then, taking into account various theoretical and experimental constraints, including those
from low-energy measurements and the Planck data on the relic DM density, we scan
the model parameter space for regions that can accommodate the Higgs exotic decays
and also masses of the new particles that can be produced at $e^+e^-$ colliders.
At present the LHC Higgs data do not yet translate into severe restrictions on the allowed
parameter values.
Subsequently, we consider constraints on them from past measurements at LEP\,II on
$e^+e^-$ collisions into a pair of charged leptons plus missing energy and into a~photon
plus missing energy.
These processes, respectively, receive contributions from the scotogenic reactions
\,$e^+e^-\to H^+H^-$\, and  \,$e^+e^-\to\gamma N N',\gamma\cal SP$\, followed by the (sequential)
decays of $H$, $N^{(\prime)}$, $\cal S$, and $\cal P$ into the DM particle $N_1$ plus light leptons.
We show that the $H^+H^-$ channel is subject to strict extra limitations from the LEP\,II data,
whereas the neutral channels are~not.
Finally, we turn to the possibility of measuring the same $e^+e^-$ scattering processes
with higher energies and much improved precision at a future facility, in particular the ILC.
We find that at the ILC such experiments can be expected to offer complementary
information for probing the scotogenic model more extensively.
Needless to say, future data on the Higgs boson's properties from the ILC will also be of
great relevance to checking the model.

\acknowledgments

This research was supported in part by the MOE Academic Excellence Program (Grant No. 102R891505)
and the NCTS.

\appendix

\section{Solutions for Yukawa couplings $\bm{{\cal Y}_{rk}}$\label{yjk}}

The diagonalization relations in Eq.\,(\ref{numass}) can be exactly solved for
the three pairs of Yukawa couplings $(Y_{ek},Y_{\mu k})$, \,$k=1,2,3$,\,
in terms of \,$Y_k=Y_{\tau k}$.\,
There are in total 27 sets of the possible solutions.
One can express the pairs in each set as
\,$(Y_{ek},Y_{\mu k})=\bigl(\bar e_z^{},\bar\mu_z^{}\bigr)Y_k^{}$,\, where \,$z=a,b$, or $c$\, and
\begin{eqnarray} \label{em} & \displaystyle
\bar e_a^{} \,\,=\,\, \frac{-c_{12\,}^{}c_{13}^{}}
{c_{12\,}^{}c_{23\,}^{}s_{13\,}^{}e^{i\delta}-s_{12\,}^{}s_{23}^{}} ~, ~~~~~~~ &
\bar\mu_a^{} \,\,=\,\, \frac{c_{12\,}^{}s_{23\,}^{}s_{13\,}^{}e^{i\delta}+s_{12\,}^{}c_{23}^{}}
{c_{12\,}^{}c_{23\,}^{}s_{13\,}^{}e^{i\delta}-s_{12\,}^{}s_{23}^{}} ~,
\nonumber \\ & \displaystyle
\bar e_b^{} \,\,=\,\, \frac{-s_{12\,}^{}c_{13}^{\vphantom{\int^|}}}
{s_{12\,}^{}c_{23\,}^{}s_{13\,}^{}e^{i\delta}+c_{12\,}^{}s_{23}^{}} ~, ~~~~~~~ &
\bar\mu_b^{} \,\,=\,\, \frac{s_{12\,}^{}s_{23\,}^{}s_{13\,}^{}e^{i\delta}-c_{12\,}^{}c_{23}^{}}
{s_{12\,}^{}c_{23\,}^{}s_{13\,}^{}e^{i\delta}+c_{12\,}^{}s_{23}^{}} ~,
\nonumber \\ & \displaystyle
\bar e_c^{} \,\,=\,\, \frac{s_{13\,}^{}e^{-i\delta}}{c_{23\,}^{}c_{13}^{}} ~, \hspace{19ex} &
\bar\mu_c^{} \,\,=\,\, \frac{s_{23}^{\vphantom{\int^|}}}{c_{23}^{}} ~,
\end{eqnarray}
with \,$c_{mn}^{}=\cos\theta_{mn}^{}$\, and \,$s_{mn}^{}=\sin\theta_{mn}^{}$.\,
Not all of the solution sets are desirable and lead to at least two nonzero masses among
the eigenvalues $m_{1,2,3}^{}$ in~Eq.\,(\ref{numass}).
Particularly, three of the sets can each only give one nonzero mass,
while 18 (six) of the others can yield two (three) nonzero masses.
We remark that the form of Eq.\,(\ref{Mnu}) also appears in some other models of radiative
neutrino mass~\cite{slim,Ahriche:2013zwa,2loop}, and so these solutions for $Y_{\ell k}$ are
also applicable to those models, with $\Lambda_{1,2,3}$ hiding the model details.

\section{Cross sections and decay rates\label{csr}}

From the amplitude for \,$e^+(p_+)_{\,}e^-(p_-)\to H^+H^-$\, in
Eq.\,(\ref{Mee2HH}), we arrive at the cross section
\begin{eqnarray} \label{sigma_ee2HH}
\sigma_{e\bar e\to H\bar H}^{} &\,=\,&
\frac{\pi\,\alpha^2\beta^3}{3 s} \,+\,
\frac{\alpha}{12}\; \frac{\bigl(g_L^2+g_L^{}g_R^{}\bigr)\beta^3}{s-m_Z^2} \,+\,
\frac{\bigl(g_L^4+g_L^2g_R^2\bigr)\beta^3s}{96\pi\,\bigl(s-m_Z^2\bigr)\raisebox{1.7pt}{$^{\!2}$}}
\,+\,
\raisebox{0.5ex}{\footnotesize$\displaystyle\sum_k$}\, \frac{|{\cal Y}_{1k}|^4}{64\pi\,s}
\Biggl(w_k^{}\,\ln\frac{w_k^{}+\beta}{w_k^{}-\beta}-2\beta\Biggr)
\nonumber \\ && \! +\;
\Biggl[ \frac{\alpha}{16 s} + \frac{g_L^2}{64\pi\,\bigl(s-m_Z^2\bigr)} \Biggr]
\raisebox{0.5ex}{\footnotesize$\displaystyle\sum_k$}\,|{\cal Y}_{1k}|^2
\biggl[\bigl(w_k^2-\beta^2\bigr)\,\ln\frac{w_k^{}+\beta}{w_k^{}-\beta}-2\beta\,w_k^{}\biggr]
\nonumber \\ && \! +\;
\raisebox{0.5ex}{\footnotesize$\displaystyle\sum_{j,\,k>j}$}\,
\frac{|{\cal Y}_{1j\,}{\cal Y}_{1k}|^2}{64\pi\,s}
\Biggl( \frac{w_j^2-\beta^2}{w_j^{}-w_k^{}}~\ln\frac{w_j^{}+\beta}{w_j^{}-\beta}
+ \frac{w_k^2-\beta^2}{w_k^{}-w_j^{}}~\ln\frac{w_k^{}+\beta}{w_k^{}-\beta} - 2\beta \Biggr) \,,
\end{eqnarray}
where \,$s=(p_++p_-)^2$,\, we have assumed that $s$ is not close to the $Z$ pole, \,$j,k=1,2,3$,\,
\begin{eqnarray}
\alpha \,\,=\,\, \frac{e^2}{4\pi} ~, \hspace{7ex}
\beta \,\,=\,\, \sqrt{1-\frac{4m_H^2}{s}} ~, \hspace{7ex}
w_k^{} \,\,=\,\, 1 \,+\, \frac{2M_k^2}{s} \,-\, \frac{2m_H^2}{s} \,\,>\,\, \beta ~.
\end{eqnarray}
We always take $e^\pm$ to be massless in our treatment of their scattering.
The form of $\sigma_{e\bar e\to H\bar H}$ due to the $\gamma$- and $Z$-exchange diagrams
plus a third contribution mediated by only one $N$ has been known before in
the literature~\cite{Komamiya:1988rs}.
In numerical computation of the $e^+e^-$ collisions, we employ the effective values
\,$\alpha=1/128$,\, $g=0.6517$,\, and \,$s_{\rm w}^2=0.23146$\,~\cite{pdg}.
It is worth noting that in our examples of $\sigma_{e\bar e\to H\bar H}$
the $N_k$-mediated contributions tend to dominate the $\gamma$ and $Z$ diagrams,
except in several instances where the different contributions are roughly
comparable in size.

The neutral counterpart of the preceding transition is \,$e^+e^-\to\cal SP$,\, but it is
generated at tree level by only one $Z$-exchange diagram.
Its cross section is
\begin{eqnarray}
\sigma_{e\bar e\to\cal S P}^{} \,\,=\,\,
\frac{\bigl(g_L^{}-g_R^{}\bigr)\raisebox{1pt}{$^2$}\bigl(g_L^2+g_R^2\bigr)
\bigl(s-4m_0^2\bigr)\raisebox{1pt}{$^{3/2}$}}
{96\pi\,\sqrt s\,\bigl(s-m_Z^2\bigr)\raisebox{1pt}{$^2$}}
\end{eqnarray}
for \,$m_{\cal S}^{}\simeq m_{\cal P}^{}\simeq m_0^{}$\, and $s$ away from the $Z$ pole.

For the particle masses in our illustrations, $H^\pm$ and their neutral partners, $\cal S$ and
$\cal P$, decay predominantly into the two-body final states \,$\ell_r^\pm N_k^{}$\, and
\,$\nu N_l^{}$,\, respectively, if kinematically permitted.
From~Eq.\,(\ref{LN}), we acquire their rates to be
\begin{eqnarray} \label{H2lN}
\Gamma_{H\to\ell_r N_k} &\,=\,&
\frac{|{\cal Y}_{rk}|^2}{16\pi_{\,}m_H^3} \bigl(m_H^2-m_{\ell_r}^2-M_k^2\bigr)
\sqrt{\bigl(m_H^2-m_{\ell_r}^2-M_k^2\bigr)\raisebox{1.7pt}{$^{\!2}$}-4m_{\ell_r}^2 M_k^2} ~,
\\ \label{S2nN}
\Gamma_{\hat\eta\to\nu N_l^{}} &\,=\,& \raisebox{2pt}{\footnotesize$\displaystyle\sum_r$}\,
\frac{|{\cal Y}_{rl}|^2}{16\pi_{\,}m_{\hat\eta}^3}
\bigl(m_{\hat\eta}^2-M_l^2\bigr)\raisebox{1.7pt}{$^{\!2}$} ~, ~~~~~~~
\hat\eta \,\,=\,\, \cal S {\rm~or}~ P ~.
\end{eqnarray}
Therefore, for the total widths of $H^\pm$ and $\hat\eta$, we make the approximations
\,$\Gamma_H=\mbox{\small$\sum_{\scriptscriptstyle r,k\,}$}\Gamma_{H\to\ell_r N_k}$\, and
\,$\Gamma_{\hat\eta}=\mbox{\small$\sum_{\scriptscriptstyle k\,}$}\Gamma_{\hat\eta\to\nu N_k}$\,
in our computation.

For the decays of $N_k$ if \,$k=2,3$,\, the two-body modes
\,$N_k^{}\to\ell_{r\,}^\pm H^\mp$\, and \,$N_k^{}\to\nu\hat\eta$\,
may take place with rates
\begin{eqnarray} \label{N2lH}
\Gamma_{N_k^{}\to\ell_r^+H^-} \,\,=\,\, \Gamma_{N_k^{}\to\ell_r^-H^+} &\,=\,&
\frac{|{\cal Y}_{rk}|^2\bigl(M_k^2+m_{\ell_r}^2-m_H^2\bigr)}{32\pi_{\,}M_k^3}
\sqrt{\bigl(M_k^2-m_{\ell_r}^2-m_H^2\bigr)\raisebox{1.7pt}{$^{\!2}$}-4m_{\ell_r}^2 m_H^2} ~,
\nonumber \\
\Gamma_{N_k^{}\to\nu\hat\eta}^{} &\,=\,& \raisebox{2pt}{\footnotesize$\displaystyle\sum_r$}\,
\frac{|{\cal Y}_{rk}|^2}{32\pi_{\,}M_k^3}\bigl(M_k^2-m_{\hat\eta}^2\bigr)^2 ~, ~~~~~~~
\hat\eta \,\,=\,\, \cal S {\rm~or}~ P ~.
\end{eqnarray}
If these channels are closed, $N_k$ will instead undergo \,$N_k^{}\to\nu_o^{}\nu_r^{}N_l^{}$
and possibly \,$N_k^{}\to\ell_o^-\ell_r^+N_l^{}$,\, mediated by $\hat\eta$ and $H^\pm$,
respectively.
They lead to the combined rates
\begin{eqnarray} \label{N2ffN}
\Gamma_{N_k^{}\to\nu\nu'N_l^{}} \,\,=\,\, \mbox{$\frac{1}{2}$}\;
\raisebox{2pt}{\footnotesize$\displaystyle\sum_{o,r}$}\,
\Gamma_{N_k^{}\to\nu_o^{}\nu_r^{}N_l^{}} ~, \hspace{7ex}
\Gamma_{N_k^{}\to\ell\bar\ell'N_l} \,\,=\,\,
\raisebox{2pt}{\footnotesize$\displaystyle\sum_{o,r}$}\,
\Gamma_{N_k^{}\to\ell_o^{}\bar\ell_r^{}N_l^{}} ~,
\end{eqnarray}
where the factor of $\frac{1}{2}$ removes double counting of contributions with \,$o\neq r$\,
and accounts for identical Majorana neutrinos in final states with \,$o=r$.\,
The terms in these sums are of the form
\begin{eqnarray}
\Gamma_{N_k^{}\to f_o^{}(p_1)_{\,}f_r^{}(p_2)_{\,}N_l^{}(p_3)} \,\,=\,\,
\frac{2}{(8\pi_{\,}M_k\mbox{$)^3$}} \int d\bar s_{12}^{}\,d\bar s_{23}^{}\;
\overline{\bigl|{\cal M}_{N_k\to f_o^{}f_r^{}N_l}\bigr|^2} ~,
\end{eqnarray}
where \,$\bar s_{ik}^{}=(p_i+p_k)^2$\, and the expressions for the integrand are derived
in the next paragraph.
For the new particles' coupling and mass values which we have considered,
these two- and/or three-body decay modes of $N_k$ dominate its total width $\Gamma_{N_k}$.

Since $\nu$ and $N$ are Majorana fermions, from Eq.\,(\ref{LN}) the amplitude for
\,$N_k^{}\to\nu_o^{}(p_1)\,\nu_r^{}(p_2)\,N_l^{}(p_3)$\, with \,$M_k^{}<m_{\cal S,P}^{}$\, is
\begin{eqnarray} \label{M_N2nnN}
{\cal M}_{N_k^{}\to\nu_o^{}\nu_r^{}N_l^{}}^{} &\,=\,&
\frac{-\bar u_{\nu_o^{}}^{}\bigl({\cal Y}_{ok}^{}P_R^{}+{\cal Y}_{ok}^*P_L^{}\bigr)u_{N_k}^{}\,
\bar u_{\nu_r^{}}^{}\bigl({\cal Y}_{rl}^{}P_R^{}+{\cal Y}_{rl}^*P_L^{}\bigr)v_{N_l}^{}}
{2_{\,}\hat{\textsf{s}}_{23}^{m_{\cal S}}}
\nonumber \\ && \! + ~
\frac{\bar u_{\nu_o^{}}^{}\bigl({\cal Y}_{ok}^{}P_R^{}-{\cal Y}_{ok}^*P_L^{}\bigr)u_{N_k}^{}\,
\bar u_{\nu_r^{}}^{}\bigl({\cal Y}_{rl}^{}P_R^{}-{\cal Y}_{rl}^*P_L^{}\bigr)v_{N_l}^{}}
{2_{\,}\hat{\textsf{s}}_{23}^{m_{\cal P}}}
\nonumber \\ && \! + ~
\frac{\bar u_{\nu_r^{}}^{}\bigl({\cal Y}_{rk}^{}P_R^{}+{\cal Y}_{rk}^*P_L^{}\bigr)u_{N_k}^{}\,
\bar u_{\nu_o^{}}^{}\bigl({\cal Y}_{ol}^{}P_R^{}+{\cal Y}_{ol}^*P_L^{}\bigr)v_{N_l}^{}}
{2_{\,}\hat{\textsf{s}}_{13}^{m_{\cal S}}}
\nonumber \\ && \! - ~
\frac{\bar u_{\nu_r^{}}^{}\bigl({\cal Y}_{rk}^{}P_R^{}-{\cal Y}_{rk}^*P_L^{}\bigr)u_{N_k}^{}\,
\bar u_{\nu_o^{}}^{}\bigl({\cal Y}_{ol}^{}P_R^{}-{\cal Y}_{ol}^*P_L^{}\bigr)v_{N_l}^{}}
{2_{\,}\hat{\textsf{s}}_{13}^{m_{\cal P}}} ~,
\end{eqnarray}
where
\begin{eqnarray}
\hat{\textsf{s}}_{ik}^{m} \,\,=\,\, \bar s_{ik}^{}-m^2 ~, \hspace{7ex}
\bar s_{ik}^{} \,\,=\,\, (p_i+p_k)^2 ~.
\end{eqnarray}
Averaging (summing) the absolute square of this amplitude over initial (final) spins,
we then get
\begin{eqnarray} \label{M2_N2nnN}
\overline{\bigl|{\cal M}_{N_k\to\nu_o\nu_r N_l}\bigr|^2} &=&
-|{\cal Y}_{ok}{\cal Y}_{rl}|^2\,
\frac{\hat{\textsf{s}}_{23}^{M_k\,}\hat{\textsf{s}}_{23}^{M_l}}{2}
\Biggl[ \frac{1}{\bigl(\hat{\textsf{s}}_{23}^{m_{\cal S}^{}}\bigr)\raisebox{1.7pt}{$^{\!2}$}}
+ \frac{1}{\bigl(\hat{\textsf{s}}_{23}^{m_{\cal P}^{}}\bigr)\raisebox{1.7pt}{$^{\!2}$}} \Biggr]
\,-\,
|{\cal Y}_{rk}{\cal Y}_{ol}|^2\,
\frac{\hat{\textsf{s}}_{13}^{M_k\,}\hat{\textsf{s}}_{13}^{M_l}}{2}
\Biggl[ \frac{1}{\bigl(\hat{\textsf{s}}_{13}^{m_{\cal S}^{}}\bigr)\raisebox{1.7pt}{$^{\!2}$}}
+ \frac{1}{\bigl(\hat{\textsf{s}}_{13}^{m_{\cal P}^{}}\bigr)\raisebox{1.7pt}{$^{\!2}$}} \Biggr]
\nonumber \\ && \! +~
{\rm Re}\bigl({\cal Y}_{ok}^*{\cal Y}_{rl}^{}{\cal Y}_{rk}^*{\cal Y}_{ol}^{}\bigr)\,
\frac{M_k^{}M_l^{}\,\bar s_{12}^{}}{2}
\Biggl( \frac{1}{\hat{\textsf{s}}_{23}^{m_{\cal S}^{}}}
+ \frac{1}{\hat{\textsf{s}}_{23}^{m_{\cal P}^{}}} \Biggr)
\Biggl( \frac{1}{\hat{\textsf{s}}_{13}^{m_{\cal S}^{}}}
+ \frac{1}{\hat{\textsf{s}}_{13}^{m_{\cal P}^{}}} \Biggr)
\nonumber \\ && \! +~
{\rm Re}\bigl({\cal Y}_{ok}^*{\cal Y}_{rl}^*{\cal Y}_{rk}^{}{\cal Y}_{ol}^{}\bigr)\,
\frac{M_k^2 M_l^2-\bar s_{23}^{}\bar s_{13}^{}}{2}
\Biggl( \frac{1}{\hat{\textsf{s}}_{23}^{m_{\cal S}^{}}}
- \frac{1}{\hat{\textsf{s}}_{23}^{m_{\cal P}^{}}} \Biggr)
\Biggl( \frac{1}{\hat{\textsf{s}}_{13}^{m_{\cal S}^{}}}
- \frac{1}{\hat{\textsf{s}}_{13}^{m_{\cal P}^{}}} \Biggr) \,.
\end{eqnarray}
Similarly, the amplitude for \,$N_k\to\ell_o^-(p_1)\,\ell_r^+(p_2)\,N_l^{}(p_3)$\,
with \,$M_k^{}<m_H^{}+m_{\ell_o,\ell_r}$\, is
\begin{eqnarray} \label{M_N2llN}
{\cal M}_{N_k^{}\to\ell_o^{}\bar\ell_r^{}N_l^{}}^{} \,\,=\,\,
\frac{{\cal Y}_{ok}^{}{\cal Y}_{rl}^*\,\bar u_{\ell_o}^{}P_R^{}u_{N_k}^{}\,
\bar u_{N_l}^{}P_L^{}v_{\ell_r}^{}}{\hat{\textsf{s}}_{23}^{m_H}}
\,-\,
\frac{{\cal Y}_{rk}^*{\cal Y}_{ol}^{}\,\bar u_{N_l}^{}\gamma^\lambda P_L^{} u_{N_k}^{}\,
\bar u_{\ell_o}^{}\gamma_\lambda^{}P_L^{}v_{\ell_r}^{}}{2_{\,}\hat{\textsf{s}}_{13}^{m_H}} ~,
\end{eqnarray}
leading to
\begin{eqnarray} \label{M2_N2llN}
\overline{\bigl|{\cal M}_{N_k\to\ell_o\bar\ell_r N_l}\bigr|^2} &\,=\,&
-|{\cal Y}_{ok}{\cal Y}_{rl}|^2\,\frac{\bigl(\hat{\textsf{s}}_{23}^{M_k}-m_{\ell_o}^2\bigr)
\bigl(\hat{\textsf{s}}_{23}^{M_l}-m_{\ell_r}^2\bigr)}
{2\bigl(\hat{\textsf{s}}_{23}^{m_H}\bigr)\raisebox{1.7pt}{$^{\!2}$}}
\nonumber \\ && \! -~
|{\cal Y}_{rk}{\cal Y}_{ol}|^2\,\frac{\bigl(\hat{\textsf{s}}_{13}^{M_k}-m_{\ell_r}^2\bigr)
\bigl(\hat{\textsf{s}}_{13}^{M_l}-m_{\ell_o}^2\bigr)}
{2\bigl(\hat{\textsf{s}}_{13}^{m_H}\bigr)\raisebox{1.7pt}{$^{\!2}$}}
\nonumber \\ && \! +~
{\rm Re}\bigl({\cal Y}_{ok}^*{\cal Y}_{rl}^{}{\cal Y}_{rk}^*{\cal Y}_{ol}^{}\bigr)\,
\frac{M_k^{}M_l^{}\,\bigl(\bar s_{12}^{}-m_{\ell_o}^2-m_{\ell_r}^2\bigr)}
{\hat{\textsf{s}}_{23}^{m_H\,} \hat{\textsf{s}}_{13}^{m_H}} ~.
\end{eqnarray}
In the case of \,$m_0^{}\simeq m_{\cal S}^{}\simeq m_{\cal P}^{}$,\, the formulas in this
paragraph are related by crossing symmetry to those for
\,$N_k^{} N_l^{}\to\nu_o^{}\nu_r^{}$\, and \,$N_k^{} N_l^{}\to\ell_o^-\ell_r^+$\, given in
Ref.\,\cite{Ho:2013hia}.\footnote{The expression for
\,$\overline{|{\cal M}_{N_k N_l\to\nu_i\nu_j}|^2}$\, in Eq.\,(B3) of Ref.\,\cite{Ho:2013hia}
needs to be multiplied by an overall factor of 2 due to $\nu_{i,j}^{}$ being Majorana particles.
Since the final neutrinos are not observed, the corresponding cross-section
is~\,$\sigma_{N_k N_l\to\nu\nu'}=(1/2)\raisebox{1pt}{$\scriptstyle\sum$}_{i,j\,}^{}
\sigma_{N_k N_l\to\nu_i\nu_j}$,\,
where the factor of 1/2 removes double counting of contributions with \,$i\neq j$\,
and accounts for identical neutrinos in final states with \,$i=j$.\,
As a consequence, the results in Ref.\,\cite{Ho:2013hia} for the DM annihilation
\,$N_1 N_1\to\nu\nu',\ell\bar\ell'$\, are numerically unaffected.}

For the scattering \,$e^+(p_+)\,e^-(p_-)\to\gamma(\textsc{k})\,N_k^{}(q_-)\,N_l^{}(q_+)$,\,
one can define the Lorentz-invariant kinematical variables
\begin{eqnarray} \label{stuk} &
s \,\,=\,\, (p_++p_-)^2 ~, \hspace{7ex} s' \,\,=\,\, (q_++q_-)^2 ~,
& \nonumber \\ &
t \,\,=\,\, (p_+-q_+)^2 ~, \hspace{7ex} t' \,\,=\,\, (p_--q_-)^2 ~,
& \nonumber \\ &
u \,\,=\,\, (p_+-q_-)^2 ~, \hspace{7ex} u' \,\,=\,\, (p_--q_+)^2 ~,
& \nonumber \\ &
\kappa_\pm^{} \,\,=\,\, 2_{\,}\textsc{k}\cdot p_\pm^{} ~, \hspace{7ex}
\kappa_\pm' \,\,=\,\, 2_{\,}\textsc{k}\cdot q_\pm^{} ~, &
\end{eqnarray}
before deriving its amplitude ${\cal M}_{e\bar e\to\gamma N_k N_l}$.
Because of the Majorana nature of $N_{k,l}$, at tree level the amplitude comes from six diagrams
mediated by $H$ with the photon radiated from the $e^\pm$ legs and the $H$ lines.
We write it as
\begin{eqnarray} \label{ee2gNN}
{\cal M}_{e\bar e\to\gamma N_k N_l} &\,=\,&
e {\cal Y}_{1k\,}^*{\cal Y}_{1l}^{} \left\{
\frac{\bar u_{N_k}^{}P_L^{} \bigl(\not{\!p}_-^{}-\!\not{\!\textsc{k}}\bigr)
\!\!\not{\hspace{-1.2pt}\varepsilon}^*u_{e^-}^{}\,\bar v_{e^+}^{}P_{R\,}^{}v_{N_l}^{}}
{\bigl(t-m_H^2\bigr)\,\kappa_-^{}}
\right.
\nonumber \\ && \hspace{53pt} - \left.
\frac{\bar u_{N_k}^{}P_L^{}u_{e^-\,}^{}\bar v_{e^+}^{}\!\!\not{\hspace{-1.2pt}\varepsilon}^*
\bigl(\not{\!p}_+^{}-\!\not{\!\textsc{k}}\bigr)P_{R\,}^{}v_{N_l}^{}}
{\bigl(t'-m_H^2\bigr)\,\kappa_+^{}}
\right.
\nonumber \\ && \hspace{53pt} - \left.
\frac{2\bar u_{N_k}^{}P_L^{}u_{e^-\,}^{}\bar v_{e^+}^{}P_{R\,}^{}v_{N_l}^{}\,
\varepsilon^*\!\cdot\!(p_--q_-)}{\bigl(t-m_H^2\bigr)\bigl(t'-m_H^2\bigr)}
\right\}
\nonumber \\ && \! +\;
e {\cal Y}_{1k\,}^{}{\cal Y}_{1l}^* \left\{ \frac{\bar u_{N_k}^{}\gamma^\rho P_L^{}v_{N_l\,}^{}
\bar v_{e^+}^{}\gamma_\rho^{}P_L^{}\bigl(\not{\!p}_-^{}-\!\not{\!\textsc{k}}\bigr)
\!\!\not{\hspace{-1.2pt}\varepsilon}^*u_{e^-}^{}}{2\bigl(m_H^2-u\bigr)\,\kappa_-^{}}
\right.
\nonumber \\ && \hspace{63pt} - \left.
\frac{\bar u_{N_k}^{}\gamma^\rho P_L^{}v_{\bar N\,}^{}\bar v_{e^+}^{}
\!\!\not{\hspace{-1.2pt}\varepsilon}^*\bigl(\not{\!p}_+^{}-\!\not{\!\textsc{k}}\bigr)
\gamma_\rho^{}P_L^{}u_{e^-}^{}}{2\bigl(m_H^2-u'\bigr)\,\kappa_+^{}}
\right.
\nonumber \\ && \hspace{63pt} + \left.
\frac{\bar u_{N_k}^{}\gamma^\rho P_L^{}v_{N_l\,}^{}
\bar v_{e^+}^{}\gamma_\rho^{}P_L^{}u_{e^-}^{}\,\varepsilon^*\!\cdot\!(p_--q_+)}
{\bigl(u-m_H^2\bigr)\bigl(u'-m_H^2\bigr)}
\right\} .
\end{eqnarray}
It is straightforward to check that this amplitude respects electromagnetic gauge invariance.
Averaging (summing) the absolute square of \,${\cal M}_{e\bar e\to\gamma N_k N_l}$\,
over the initial (final) spins, one then obtains
\begin{eqnarray} && \hspace{-2ex}
\frac{\overline{|{\cal M}_{e\bar e\to\gamma N_k N_l}|^2}}{e^2} \,\,=\,\,
\nonumber \\ &&
\frac{|{\cal Y}_{1k}{\cal Y}_{1l}|^2}{2\kappa_-^{}} \left\{
\frac{M_l^2-t}{\bigl(m_H^2-t\bigr)^{\!2}_{\vphantom{|}}} \Biggl[
\kappa_-' + \frac{\bigl(M_k^2+t'\bigr)\kappa_-^{}
+ 2\bigl(M_k^2-t\bigr)\bigl(M_k^2-t'\bigr)}{m_H^2-t'}
- \frac{\bigl(M_k^2-t'\bigr)\bigl(t+t'\bigr)\kappa_-^{}}
{2\bigl(m_H^2-t'\bigr)\raisebox{1.7pt}{$^{\!2}$}} \Biggr] \right.
\nonumber \\ && \hspace{5ex} + ~
\frac{M_k^2-u}{\bigl(m_H^2-u\bigr)\raisebox{1.7pt}{$^{\!2}$}} \Biggl[
\kappa_+' + \frac{\bigl(M_l^2+u'\bigr)\kappa_-^{}
+ 2\bigl(M_l^2-u\bigr)\bigl(M_l^2-u'\bigr)}{m_H^2-u'}
- \frac{\bigl(M_l^2-u'\bigr)\bigl(u+u'\bigr)\kappa_-^{}}
{2\bigl(m_H^2-u'\bigr)\raisebox{1.7pt}{$^{\!2}$}} \Biggr]_{\vphantom{|_|^|}}
\nonumber \\ && \hspace{5ex} + ~
\frac{\bigl(M_l^2-t\bigr)
\bigl[\bigl(t'-s-u\bigr)\kappa_-^{}+\bigl(M_k^2-t\bigr)s\bigr]
+ \bigl(M_k^2-t'\bigr)
\bigl[\bigl(t-s-u'\bigr)\kappa_+^{}+\bigl(M_l^2-t'\bigr)s\bigr]}
{2 \bigl(m_H^2-t\bigr)\bigl(m_H^2-t'\bigr)\kappa_+^{}}
\nonumber \\ && \hspace{5ex} + ~ \left.
\frac{\bigl(M_k^2-u\bigr)
\bigl[\bigl(u'-s-t\bigr)\kappa_-^{}+\bigl(M_l^2-u\bigr)s\bigr]
+ \bigl(M_l^2-u'\bigr) \bigl[\bigl(u-s-t'\bigr)\kappa_+^{}
+ \bigl(M_k^2-u'\bigr)s\bigr]^{\vphantom{|^|}}}
{2 \bigl(m_H^2-u\bigr)\bigl(m_H^2-u'\bigr)\kappa_+^{}} \right\}
\nonumber \\ && - ~
\frac{M_k^{}M_l^{}\;{\rm Re}\bigl({\cal Y}_{1k}^{*2}{\cal Y}_{1l}^2\bigr)}
{\bigl(m_H^2-t\bigr)\bigl(m_H^2-u\bigr)\kappa_-^{}}
\Biggl[
\kappa_+^{} \,+\, \frac{\bigl(m_H^2-u\bigr)ss'}{\bigl(m_H^2-u'\bigr)\kappa_+^{}}
\,+\,
\frac{\bigl(2M_k^2+2M_l^2+s-s'\bigr)\kappa_{-\,}^{}s}
{4\bigl(m_H^2-t'\bigr)\bigl(m_H^2-u'\bigr)}
\nonumber \\ && \hspace{27ex} + ~
\frac{\bigl(M_k^2-u\bigr)\kappa_-^{} - \bigl(M_k^2-t'\bigr)\kappa_+^{}
+ (2M_k^2-t-t'\bigr)s}{2\bigl(m_H^2-t'\bigr)}
\nonumber \\ && \hspace{27ex} + ~
\frac{\bigl(M_l^2-t\bigr)\kappa_-^{} - \bigl(M_l^2-u'\bigr)\kappa_+^{}
+ (2M_l^2-u-u'\bigr)s}{2\bigl(m_H^2-u'\bigr)} \Biggr]
\nonumber \\ && + ~
\frac{4 M_k^{}M_l^{}\; {\rm Re}\bigl({\cal Y}_{1k}^*{\cal Y}_{1l}^{}\bigr)\,
{\rm Im}\bigl({\cal Y}_{1k\,}^*{\cal Y}_{1l}^{}\bigr)}
{\bigl(m_H^2-t\bigr)\bigl(m_H^2-u\bigr)\kappa_-^{}}
\Biggl(\frac{1}{m_H^2-t'}+\frac{1}{m_H^2-u'}\Biggr)
\epsilon_{\rho\sigma\tau\omega}^{}\, p_+^\rho p_-^\sigma q_+^\tau q_-^\omega
\nonumber \\ && + ~
\Bigl( t\leftrightarrow t',\, u\leftrightarrow u',\,
\kappa_-^{}\leftrightarrow\kappa_+^{},\, \kappa_-'\leftrightarrow\kappa_+',\,
M_k^{}\leftrightarrow M_l^{},\, {\cal Y}_{1k}^{}\leftrightarrow{\cal Y}_{1l}^{} \Bigr) ~.
\end{eqnarray}
This leads to the cross section
\begin{eqnarray}
\sigma_{e\bar e\to\gamma N_k N_l}^{} = \int
\frac{E_\gamma\,dE_\gamma\,d(\cos\theta_\gamma)\;d\bar\Omega_N}{2(1+\delta_{kl})(4\pi)^4\,s}
\sqrt{1-\frac{2M_k^2+2M_l^2}{s-2E_\gamma\sqrt s}
+ \Biggl(\frac{M_k^2-M_l^2}{s-2E_\gamma\sqrt s}\Biggr)\raisebox{12pt}{$^{\!\!\!2}$}} ~
\overline{|{\cal M}_{e\bar e\to\gamma N_k N_l}|^2} ~, ~~~~
\end{eqnarray}
where $E_\gamma^{}$ and $\theta_\gamma^{}$ are the photon energy and angle with respect to
the $e^+$ or $e^-$ beam direction in the c.m. frame of the $e^+e^-$ pair,
$\bar\Omega_N^{}$ denotes the solid angle of either $N_k$ or $N_l$ in the c.m.
frame of the $N_k N_l$ pair, and the factor \,$1/(1+\delta_{kl})$\, accounts for the identical
Majorana fermions in the final states with \,$k=l$.\,
The range of the photon energy is
\begin{eqnarray} \label{Egrange}
E_\gamma^{\rm min} \,\,\le\,\, E_\gamma^{}\,\,\le\,\, E_\gamma^{\rm max}
\,\,=\,\, \frac{s-(M_k+M_l)^2}{2\sqrt s}~,
\end{eqnarray}
where $E_\gamma^{\rm min}$ is an experimental cut.
In the numerical evaluation of the integral, the $\theta_\gamma^{}$ range is also subject to cuts.

For \,$e^+(p_+)\,e^-(p_-)\to\gamma(\textsc{k})\,{\cal S}(q_-)\,{\cal P}(q_+)$,\,
the kinematical variables are the same as those listed in~Eq.\,(\ref{stuk}).
This reaction is induced at tree level by two $Z$-exchange diagrams with the photon emitted
from the $e^\pm$ lines.
Its amplitude is
\begin{eqnarray} \label{ee2gSP}
{\cal M}_{e\bar e\to\gamma\cal S P} &\,=\,&
\frac{2ie\, \bar v_{e^+}^{}\!\!\not{\!q}_-^{}
\bigl[\bigl(g_L^{}g_R^{}-g_L^2\bigr)P_L^{}+\bigl(g_R^2-g_L^{}g_R^{}\bigr)P_R^{}\bigr]
\bigl(\not{\!p}_-^{}-\!\not{\!\textsc{k}}\bigr)\!\!\not{\hspace{-1.2pt}\varepsilon}^*u_{e^-}^{}}
{\bigl(s'-m_Z^2+i\Gamma_Z^{}m_Z^{}\bigr)\,\kappa_-^{}}
\nonumber \\ && \! +~
\frac{2ie\, \bar v_{e^+}^{}\!\!\not{\hspace{-1.2pt}\varepsilon}^*
\bigl(\not{\!\textsc{k}}-\!\!\not{\!p}_+^{}\bigr)\!\!\not{\!q}_-^{}
\bigl[\bigl(g_L^{}g_R^{}-g_L^2\bigr)P_L^{}+\bigl(g_R^2-g_L^{}g_R^{}\bigr)P_R^{}\bigr]u_{e^-}^{}}
{\bigl(s'-m_Z^2+i\Gamma_Z^{}m_Z^{}\bigr)\,\kappa_+^{}} ~,
\end{eqnarray}
where $g_{L,R}^{}$ are defined in Section\,\,\ref{ee2HH}.
One can easily verify that ${\cal M}_{e\bar e\to\gamma\cal S P}$ is electromagnetically
gauge invariant.
It follows that
\begin{eqnarray}
\sigma_{e\bar e\to\gamma\cal SP}^{} = \int
\frac{E_\gamma\,dE_\gamma\,d(\cos\theta_\gamma)\;d\bar\Omega_{\hat\eta}}{2(4\pi)^4\,s}
\sqrt{1-\frac{4m_0^2}{s-2E_\gamma\sqrt s}} ~~
\overline{|{\cal M}_{e\bar e\to\gamma\cal SP}|^2} ~,
\end{eqnarray}
where $\bar\Omega_{\hat\eta}^{}$ denotes the solid angle of either $\cal S$ or $\cal P$ in
the c.m. frame of the $\cal SP$ pair and
\begin{eqnarray}
\overline{|{\cal M}_{e\bar e\to\gamma\cal S P}|^2} \,\,=\,\,
2e^2\bigl(g_L^{}-g_R^{}\bigr)\raisebox{1pt}{$^2$}\bigl(g_L^2+g_R^2\bigr)\;
\frac{s'\bigl(t u+t'u'+m_0^2 s-2m_0^4\bigr)+m_0^2\bigl(2\kappa_+^{}\kappa_-^{}-s^2\bigr)}
{\kappa_+^{}\kappa_-^{~~}\bigl[\bigl(s'-m_Z^2\bigr)\raisebox{1pt}{$^2$}+\Gamma_Z^2m_Z^2\bigr]}
~. ~~~~
\end{eqnarray}
The photon energy range in this case is
\begin{eqnarray} \label{egrange}
E_\gamma^{\rm min} \,\,\le\,\, E_\gamma^{}\,\,\le\,\, E_\gamma^{\rm max}
\,\,=\,\, \frac{s-4m_0^2}{2\sqrt s} ~.
\end{eqnarray}

Lastly, it is instructive to compare our calculation of
\,$\sigma_{e\bar e\to\gamma N_k N_l,\gamma\cal SP}$\,
above with its estimation in the so-called radiator approximation~\cite{Nicrosini:1988hw}.
For \,$XY=N_k N_l$\, or \,$\cal SP$,\, it is given by
\begin{eqnarray} & \displaystyle
\sigma_{e\bar e\to\gamma XY}^{} \,\,\simeq\,\, \int dc_\gamma^{}\,dx_\gamma^{}\,
{\cal H}(c_\gamma,x_\gamma;s)\, \hat\sigma(\hat s) ~,
& \\ & \displaystyle
c_\gamma^{} \,=\, \cos\theta_\gamma^{} ~, \hspace{5ex}
x_\gamma^{} \,=\, \frac{2 E_\gamma}{\sqrt s} ~, \hspace{5ex}
{\cal H}(c,x;s) \,=\, \frac{\alpha}{\pi}\;\frac{(2-x)^2+c^2 x^2}{2(1-c^2)x} ~, \hspace{5ex}
\hat s \,=\, (1-x_\gamma)s ~, & ~~~~ \nonumber
\end{eqnarray}
where $\hat\sigma(\hat s)$ denotes the cross section of the simpler reaction \,$e^+e^-\to XY$.\,
Thus we acquire numbers which are smaller than their counterparts
in Table\,\,\ref{csee2gE} by less than~9\%.
In contrast, our application of this approximate method to
\,$\sigma_{e^+e^-\to\gamma\nu\bar\nu}^{\rm SM}$,\, with $\hat\sigma(\hat s)$ now being
the SM cross-section of \,$e^+e^-\to\nu\bar\nu$,\, works as well only for
the \,$\sqrt s=250$\,GeV\, case, its result exceeding the corresponding number in the bottom
row of Table\,\,\ref{csee2gE} by about 9\%, whereas the estimates for
\,$\sqrt s=500,1000$~GeV\, overshoot their counterparts in the table by more than 100\%.

\end{document}